\documentclass[aps,preprint,onecolumn,secnumarabic,nobalancelastpage,amsmath,amssymb,
nofootinbib]{revtex4}

\RequirePackage{fix-cm}
\usepackage{enumerate}

\usepackage{graphicx}      
\usepackage{longtable}     
\usepackage{url}           
\usepackage{dcolumn}
\usepackage{bm}            
\usepackage{mathrsfs}
\usepackage{epstopdf}
\usepackage{color}
\usepackage[usenames,dvipsnames,svgnames]{xcolor}
\begin{document}

\preprint{APS/123-QED}

\title{Analytical solutions of the geodesic equation in the spacetime of a black hole in $f(R)$ gravity}

\author{Saheb Soroushfar${}^1$}
\author{Reza Saffari${}^1$}%
\email{rsk@guilan.ac.ir}
\author{Jutta Kunz${}^2$}
 \author{Claus L\"ammerzahl${}^{2,3}$}

\affiliation{${}^1$Department of Physics, University of Guilan, 41335-1914, Rasht, Iran.\\
${}^2$Institut f\"ur Physik, Universit\"at Oldenburg, Postfach 2503 D-26111 Oldenburg, Germany.\\
${}^3$ZARM, University of Bremen, Am Fallturm, 28359 Bremen, Germany.
}%



\date{\today}

\begin{abstract}
We consider the motion of test particles
in the spacetime of a black hole in $f(R)$ gravity.
The complete set of analytic solutions of the geodesic equation
in the spacetime of this black hole are presented.
The geodesic equations are solved in terms of Weierstrass elliptic functions
and derivatives of Kleinian sigma functions.
The different types of the resulting orbits are characterized
in terms of the conserved energy and angular momentum
as well as the cosmological constant $(\Lambda)$
and the real constant $(\beta)$.
\end{abstract}

\maketitle

\section{INTRODUCTION}

Recent observations of the Supernova Type Ia (SNIa)
and Cosmic Microwave Background (CMB) radiation indicate
that the expansion of our Universe is accelerating \cite{Riess:1998cb}.
This accelerating expansion is one of the most important puzzles
of contemporary physics.
While a nonzero vacuum energy could cause the Universe to accelerate,
one is left wondering why the vacuum energy 
would be so small
\cite{Sahni:1999gb,Carroll:2000fy}.
As a consequence, all observations related to gravity
should be described within a framework including the cosmological constant.
However, due to the smallness of the cosmological constant
it seems unlikely that it will have an observable effect
on smaller, that is, on solar system scales.
In fact, it has been shown that, 
based on the Schwarzschild-de Sitter spacetime,
the cosmological constant plays no role in solar system observations
\cite{Kagramanova:2006ax}
and neither in strong fields \cite{Hackmann:2008zz}.

Another question concerning the cosmological constant is the
coincidence problem,
i.e., the fact that the energy density of matter and
of the vacuum (cosmological constant)
are on the same order of magnitude at the present time.
A model of varying dark energy can partially solve this problem.
Here the density of dark energy traces the density of matter
from the early universe to the present time.
Modified gravity can also provide an effective time varying equation of state.
In such models the Einstein-Hilbert action is replaced
with a generic form of $f(R)$ gravity \cite{Carroll:2003wy}.
But modifying the action not only affects the dynamics of the Universe,
it can also alter the dynamics at the galactic or solar system scales.

Although the majority of gravitational effects can be discussed
using approximations and numerics, a systematic study of all effects
can only be achieved by using analytical methods.
The motion of test particles (both massive and massless)
provides the only experimentally feasible way to study
the gravitational fields of objects such as black holes.
Predictions about observable effects
(such as light deflection, gravitational time-delay, perihelion shift
and the Lense-Thirring effect) can be made and compared with observations.
For this purpose the Weierstrassian elliptic functions are most useful
because they lead to simple expressions.
The resulting structure of the equations of motion
is essentially the same as in the Schwarzschild spacetime,
where they can be solved analytically in terms of elliptic functions,
as first demonstrated by Hagihara in 1931 \cite{Hagihara:1931}.

This analytical method was further advanced and applied to the hyperelliptical case,
where the analytical solution
of the equations of motion in the 4-dimensional Schwarzschild-de Sitter
\cite{Hackmann:2008zz} and Kerr-de Sitter spacetime \cite{Hackmann:2010zz}
was presented.
It was applied as well as in the higher-dimensional Schwarzschild,
Schwarzschild-(anti)de Sitter, Reissner-Nordstr\"om
and Reissner-Nordstr\"om -(anti)-de Sitter spactime \cite{Hackmann:2008tu}
and in the higher-dimensional Myers-Perry spacetime
\cite{Enolski:2010if,Kagramanova:2012hw}.
Also the general solutions to the geodesic equation in Schwarzschild
and Kerr black hole pierced by a cosmic string
have been discussed extensively \cite{Hackmann:2009rp,Hackmann:2010ir}.
Recently, the geodesic equations were solved analytically
in special cases
in the singly spinning black ring spacetime \cite{Grunau:2012ai},
as well as in the (charged) doubly spinning black ring spacetime
\cite{Grunau:2012ri} and in the (rotating) black string spacetime
\cite{Grunau:2013oca}.

In this paper we discussed the motion of test particles
in the spacetime of the black hole derived in modified gravity
\cite{Saffari:2007zt}.
We present the results here in terms of Weierstrass elliptic functions
and derivatives of Kleinian sigma functions.
Our paper is organized as follows:
in Section (\ref{field}) we give a brief review of the field equations
in $f(R)$ modified gravity.
In Section (\ref{geodesic}) we present the geodesic equations
describing test particle motion in the spacetime of this black hole
and we give our analytical solutions.
We conclude in Section (\ref{conclusions}).

\section{FIELD EQUATIONS IN $f(R)$ MODIFIED GRAVITY}\label{field}

In this section we give a brief review of the field equations in
$f(R)$ gravity.
A generic form of the action depending on the Ricci scalar
can be written as follows:
\begin{equation}\label{action}
S=\frac{1}{2k}\int d^{4}{x}\sqrt{-g}f(R)+S_{m}.
\end{equation}
Varying the action with respect to the metric results in the field equations as:
\begin{equation}
F(R)R_{\mu\nu}-\frac{1}{2}f(R)g_{\mu\nu}-(\nabla_{\mu}\nabla_{\nu}-g_{\mu\nu}\square)F(R)=kT_{\mu\nu},
\end{equation}
where $ F(R)=\frac{df(R)}{dr} $ and $ \square=\nabla_{\alpha}\nabla^{\alpha}$.

The metric of the spherically symmetric spacetime we are considering is given by \cite{Saffari:2007zt}
\begin{equation}
ds^{2}=B(r)dt^{2}-A(r)dr^{2}-r^{2}(d\theta^{2}+\sin^{2}\theta d\varphi^{2}),\quad
\end{equation},
where $A(r)=B(r)^{-1}$. The model employed for $f(R)$ gravity is
given by
\begin{equation}
f(R)=R+\Lambda+\frac{R+\Lambda}{R/R_0 +
2/\alpha}\ln\frac{R+\Lambda}{R_c},
\end{equation}
where $\Lambda$ is the cosmological constant, which has a value of
$|\Lambda|\leq 10^{-52} m^{-2}$ \cite{Hackmann:2008zz}, $R_c$ is a
constant of integration and $R_0=6{\alpha}^2/d^2$, where $\alpha$
and $d$ are free parameters of the action. The metric solution up to
the first order in the free parameters of the action is obtained as
$B(r)=1-\frac{2m}{r}+\beta r-\frac{1}{3}\Lambda{r}^{2}$, where
$\beta=\alpha/d\geq0$ is a real constant \cite{Saffari:2007zt}. In
this model $d$ is a scale factor around $10~kpc$, and $\alpha\simeq
10^{-6}$ is a dimensionless parameter which provides a flat rotation
curve for stars in a typical spiral galaxy.

In the case of strong curvatures where $R\gg\Lambda$ and
$R/R_0\gg2/\alpha$, the action reduces to
\begin{equation}
f(R)=R+R_0\ln\frac{R}{R_c},
\end{equation}
which is the desired action for the solar system, typical black
holes as well as spiral galaxies. On the other hand on cosmological
scales, where $R\simeq R_0\simeq\Lambda$, and when $\alpha\ll1$, the
action reduces to $f(R)=R+\Lambda$ which satisfies the late time
accelerating expansion of the Universe \cite{Saffari:2007zt}.

Therefore the small free parameter, which is denoted by $\beta$ in
this paper, satisfies both the small and the large scale physical
evidence for the two ranges of the scalar curvature, $R\gg\Lambda$
and $R\simeq\Lambda$.

\section{THE GEODESIC EQUATION}\label{geodesic}

We consider the geodesic equation
 \begin{equation}\label{2}
 \frac{d^2x^\mu}{ds^2}+\Gamma^\mu_{\rho\sigma}\frac{dx^\rho}{ds}\frac{dx^\sigma}{ds}=0,
\end{equation}
where
$ds^2=g_{\mu\nu}dx^\mu dx^\nu$
is the proper time along the geodesics
and
\begin{equation}
\Gamma^\mu_{\rho\sigma}=\frac{1}{2}g^{\mu\nu}(\partial_{\rho}g_{\sigma\nu}+\partial_{\sigma}g_{\rho\nu}-\partial_{\nu}g_{\rho\sigma})
\end{equation}
is the Christoffel symbol, in a spacetime given by the
metric
\begin{equation}\label{metric}
ds^{2}=(1-\frac{2m}{r}+\beta{r}-\frac{1}{3}\Lambda{r}^{2})dt^{2}-(1-\frac{2m}{r}+\beta{r}-\frac{1}{3}\Lambda{r}^{2})^{-1}dr^{2}-r^{2}(d\theta^{2}+\sin^{2}\theta d\varphi^{2}),\quad
\end{equation}
which describes the spherically symmetric vacuum solution of Eq.(\ref{action}). For a general discussion of this metric, see e.g.\cite{Saffari:2007zt}.
The Lagrangian $ \mathfrak{L} $ for a point particle in the space–time Eq.(\ref{metric}) reads :
\begin{equation}
\mathfrak{L}=\frac{1}{2}g_{\mu\nu}\frac{dx^\mu}{ds}\frac{dx^\nu}{ds}=\frac{1}{2}\epsilon=\frac{1}{2}[B(r)(\frac{dt}{ds})^{2}-B(r)^{-1}(\frac{dr}{ds})^{2}-r^{2}((\frac{d\theta}{ds})^{2}+\sin^{2}\theta(\frac{d\varphi}{ds})^{2})],
\end{equation}
where for massive particles
$ \epsilon=1 $
and for light
$ \epsilon=0 $

Because of the spherical symmetry we can restrict our
considerations to the equatorial plane. Furthermore, due to
the conserved energy and angular momentum
\begin{equation}
E=g_{tt}\frac{dt}{ds}=(1-\frac{2m}{r}+\beta{r}-\frac{1}{3}\Lambda{r}^{2})\frac{dt}{ds},
\end{equation}
\begin{equation}
L=g_{\varphi\varphi}\frac{d\varphi}{ds}=r^{2}\frac{d\varphi}{ds},
\end{equation}
the geodesic equation reduces to an ordinary differential
equation
\begin{equation}\label{dr/ds}
(\frac{dr}{ds})^2=E^2-(1-\frac{2m}{r}+\beta{r}-\frac{1}{3}\Lambda{r}^{2})(\epsilon+\frac{L^2}{r^2}).
\end{equation}

Together with energy and angular momentum conservation
we obtain the corresponding equations for $r$ as a function of
$ \phi $ and as a function of $t$
\begin{equation}\label{dr/dphi}
(\frac{dr}{d\varphi})^2=\frac{r^4}{L^2}(E^2-(1-\frac{2m}{r}+\beta{r}-\frac{1}{3}\Lambda{r}^{2})(\epsilon+\frac{L^2}{r^2}))=:R(r),
\end{equation}
\begin{equation}\label{dr/dt}
(\frac{dr}{dt})^2=\frac{1}{E^2}(1-\frac{2m}{r}+\beta{r}-\frac{1}{3}\Lambda{r}^{2})^2(E^2-(1-\frac{2m}{r}+\beta{r}-\frac{1}{3}\Lambda{r}^{2})(\epsilon+\frac{L^2}{r^2})).
\end{equation}
\\

Eqs.(\ref{dr/ds})-(\ref{dr/dt}) give a complete description of the
dynamics.
Eq.(\ref{dr/ds}) suggests the introduction of an effective
potential
\begin{equation}
V_{eff}=(1-\frac{2m}{r}+\beta{r}-\frac{1}{3}\Lambda{r}^{2})(\epsilon+\frac{L^2}{r^2}).
\end{equation}

The shape of an orbit depends on the energy $E$ and the angular momentum $L$
of the test particle or light ray under consideration
as well as the cosmological constant $ \Lambda $
and the real constant $ \beta $.
(The mass can be absorbed through a rescaling of the radial coordinate.)
Since $r$ should be real and positive,
the physically acceptable regions are given by those values of $r$,
for which $ E^{2}\geq V_{eff} $,
owing to the square on the left hand side of Eq.(\ref{dr/ds}).

The following different types of orbits can be identified
in the spacetimes described by the metric Eq.(\ref{metric})
\begin{enumerate}
\item Flyby orbits (FO): $r$ starts from $ \infty $, then approaches a periapsis
$ r = r_{p}$ and goes back to $\infty$.

\item Bound orbits (BO): $r$ oscillates between two boundary values
 $ r_{p} \leq\ {r} \leq r_{a} $
with
$ 0 < r_{p} < r_{a} < \infty $ .

\item Terminating bound orbits (TBO): $r$ starts in
$ (0, r_{a}]$ for $0 < r_{a} < \infty $
and falls into the singularity at $r = 0$.

\item Terminating escape orbits (TEO): $r$ comes from $ \infty $
and falls into the singularity at
$r = 0$.
\end{enumerate}

The four regular types of geodesic motion correspond to different arrangements
of the real and positive zeros of $ R(r)$ defining the borders of $ R(r) \geq 0 $ or, equivalently,
$ E^2 \geq V_{eff} $ .

All other types of orbits are exceptional and treated separately. They are connected
with the appearance of multiple zeros in $ R(r)$ or with parameter values which reduce
the degree of $ R(r)$. In both cases the structure of the differential Eq.(\ref{dr/dphi}) is
considerably simplified. These orbits are radial geodesics with $ L = 0 $
(i.e., $ \frac{dr}{d\varphi} = 0) $,
circular orbits with constant $r$,
orbits asymptotically approaching circular orbits,
and in the case of $ \beta = 0 $ and $ \Lambda = 0 $ parabolic orbits with $ E^{2} = 1 $ .

For the analysis of the dependence of the possible types
of orbits on the parameters of the spacetime and the test particle or light ray it is
convenient to use dimensionless quantities. Thus, we introduce

\begin{equation}
\tilde{r}=\frac{r}{m},\qquad
\mathcal{L}=\frac{m^{2}}{L^{2}},\qquad \tilde{\Lambda}=\frac{1}{3}\Lambda m^{2},\qquad \tilde{\beta}=\beta m.
\end{equation}

Eq.(\ref{dr/dphi}) can be written as
\begin{equation}\label{dr/dphin}
(\frac{d\tilde{r}}{d\varphi})^2=\epsilon\tilde{\Lambda}\mathcal{L}\tilde{r}^{6}-\epsilon\tilde{\beta}\mathcal{L}\tilde{r}^{5}+((E^{2}-\epsilon)\mathcal{L}+\tilde{\Lambda})\tilde{r}^{4}+(2\epsilon\mathcal{L}-\tilde{\beta})\tilde{r}^{3}-\tilde{r}^{2}+2\tilde{r}=R(\tilde{r}).
\end{equation}

Eq.(\ref{dr/dphin}) implies that $ R(\tilde{r}) \geq 0 $ is a necessary condition for the existence of a
geodesic and, thus, that the real and positive zeros of $ R(\tilde{r}) $ are extremal values of the geodesic motion.
In the limit $ \tilde \beta=0, \tilde\Lambda=0 $,
Eq.(\ref{dr/dphin}) reduces to the well known
Schwarzschild limit,
and in the limit  $ \tilde\beta=0 $
to the Schwarzschild-de Sitter limit.
As $ \tilde{r}= 0 $ is a zero of  $ R(\tilde{r}) $
for all values of the parameters, it is
neglected in the following analysis and
\begin{equation}
R^{*}(\tilde{r})=\epsilon\tilde{\Lambda}\mathcal{L}\tilde{r}^{5}-\epsilon\tilde{\beta}\mathcal{L}\tilde{r}^{4}+((E^{2}-\epsilon)\mathcal{L}+\tilde{\Lambda})\tilde{r}^{3}+(2\epsilon\mathcal{L}-\tilde{\beta})\tilde{r}^{2}-\tilde{r}+2
\end{equation}
is considered instead.

\begin{figure}[ht]
\centerline{\includegraphics[width=10cm]{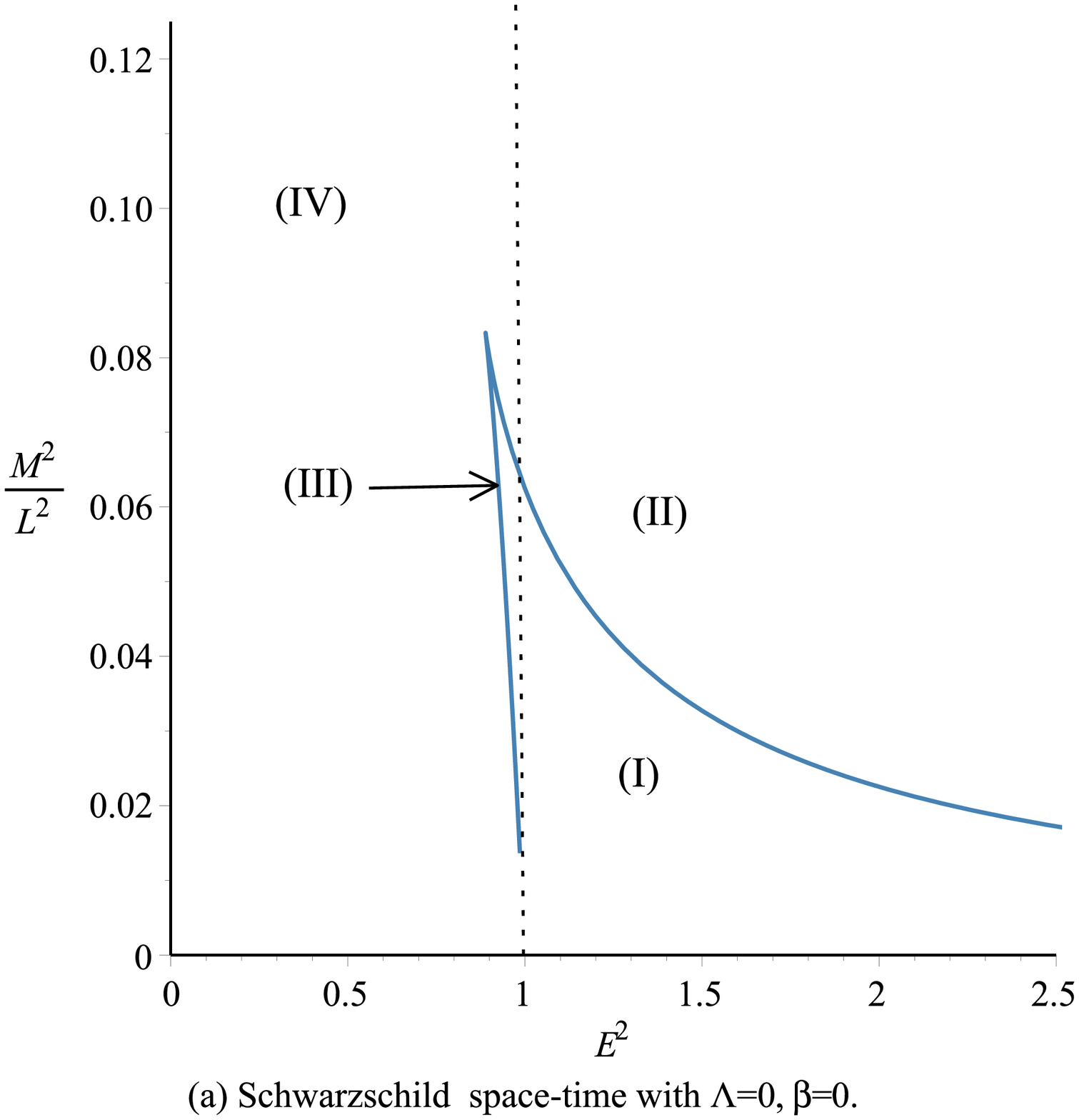}}
\centerline{\includegraphics[width=10cm]{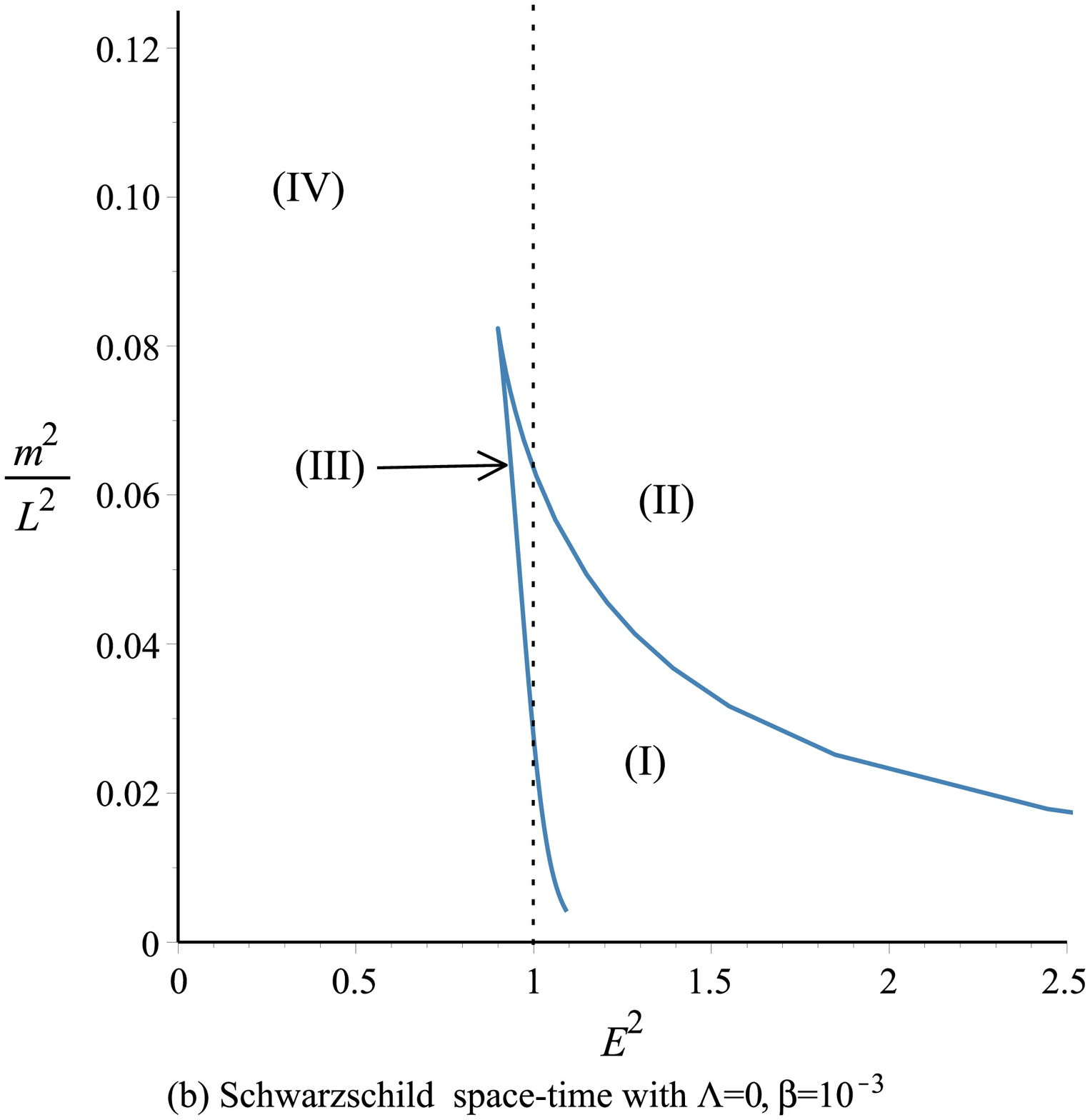}}
\caption{\label{region1}\small Regions of different types of geodesic motion for test particles ($ \epsilon=1 $).
The numbers of positive real zeros in these regions are:
(I)=2, (II)=0, (III)=3, (IV)=1.}
\end{figure}

\begin{figure}[ht]
\centerline{\includegraphics[width=10cm]{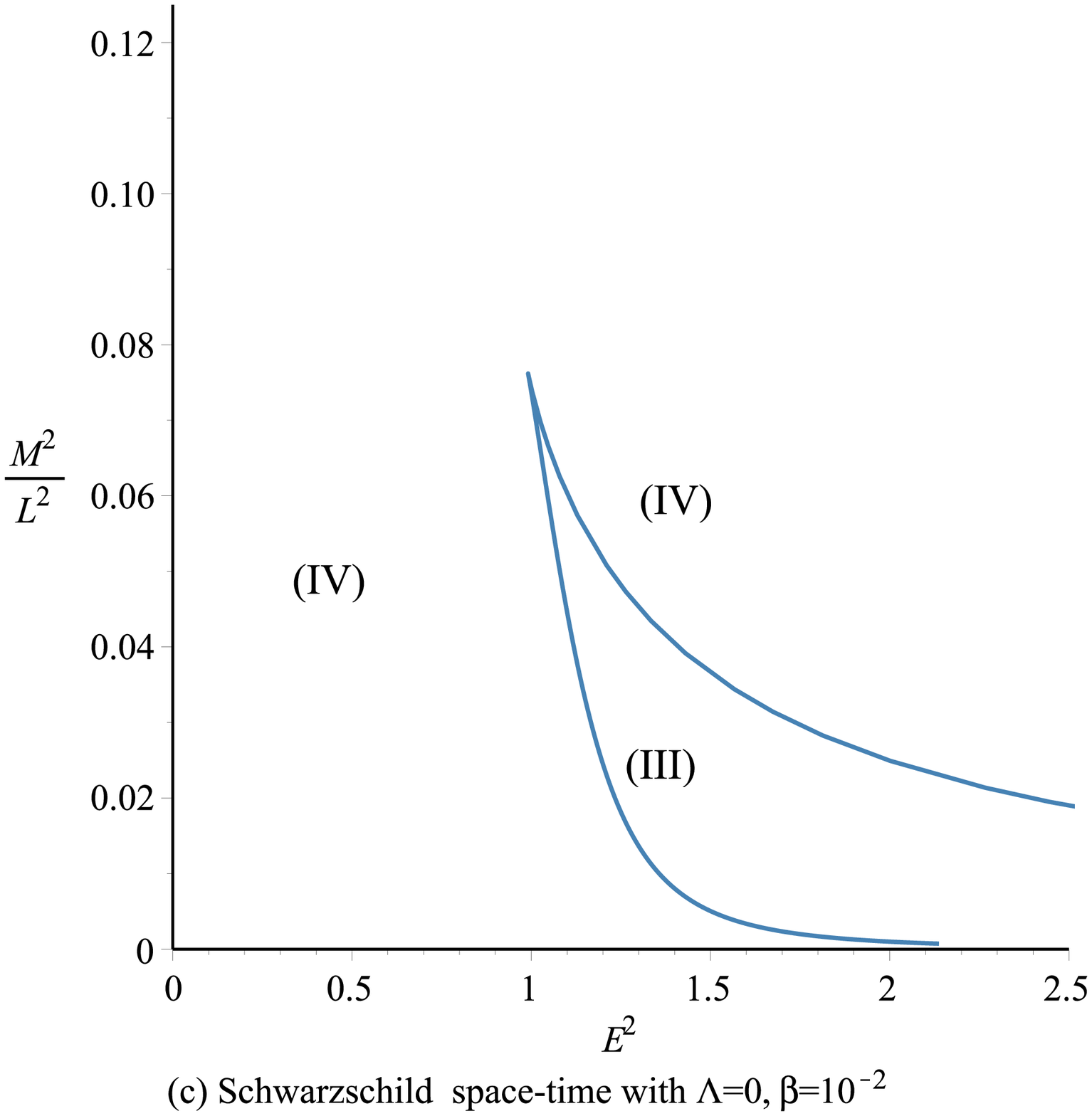}}
\centerline{\includegraphics[width=10cm]{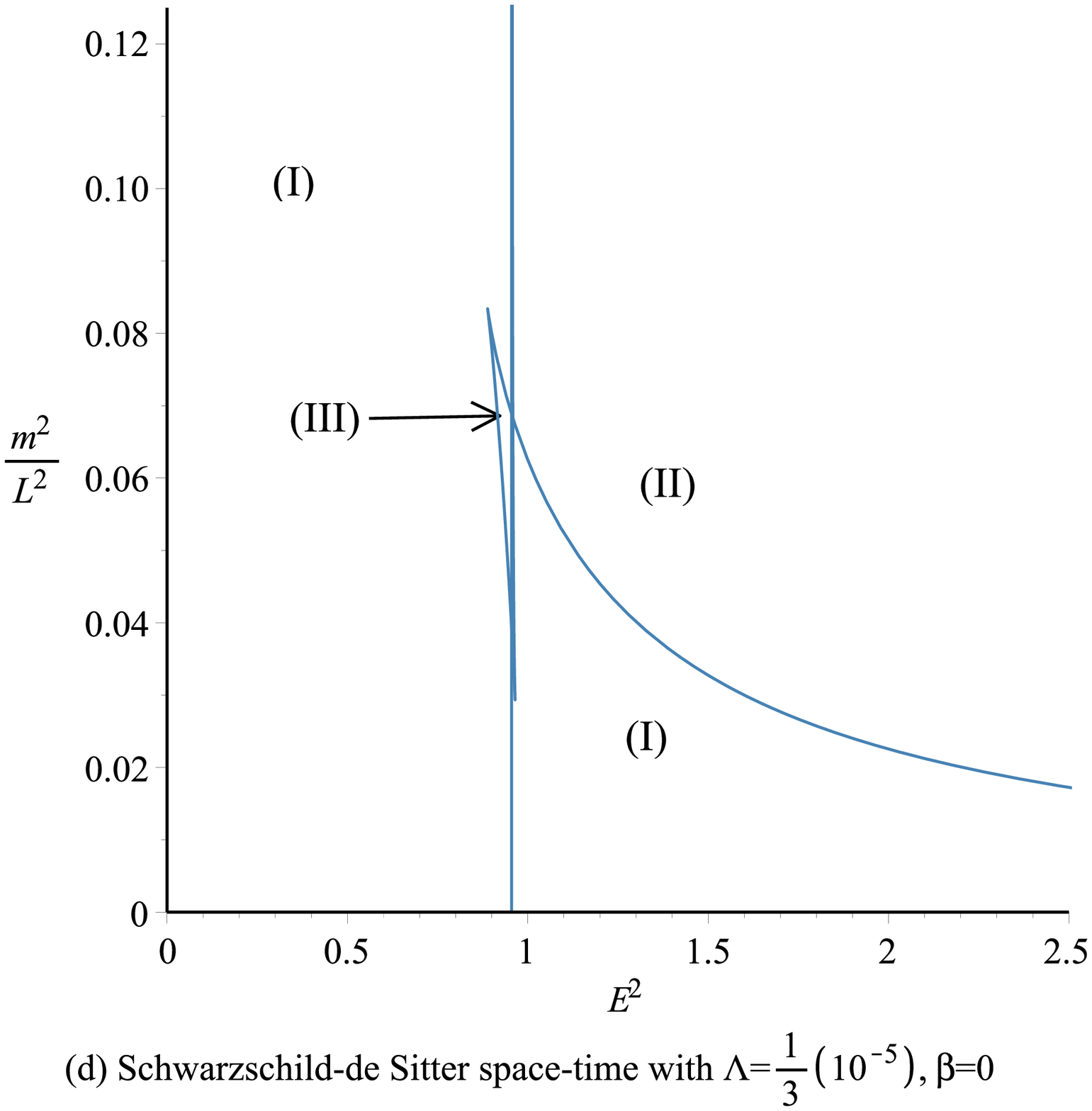}}
\caption{\label{region2}\small Regions of different types of geodesic motion for test particles ($ \epsilon=1 $). The numbers of positive real zeros in these
 regions are: (I)=2, (II)=0, (III)=3 for (Schwarzschild) and 4 for (Schwarzschild-de Sitter), (IV)=1.}
\end{figure}

\begin{figure}[ht]
\centerline{\includegraphics[width=10cm]{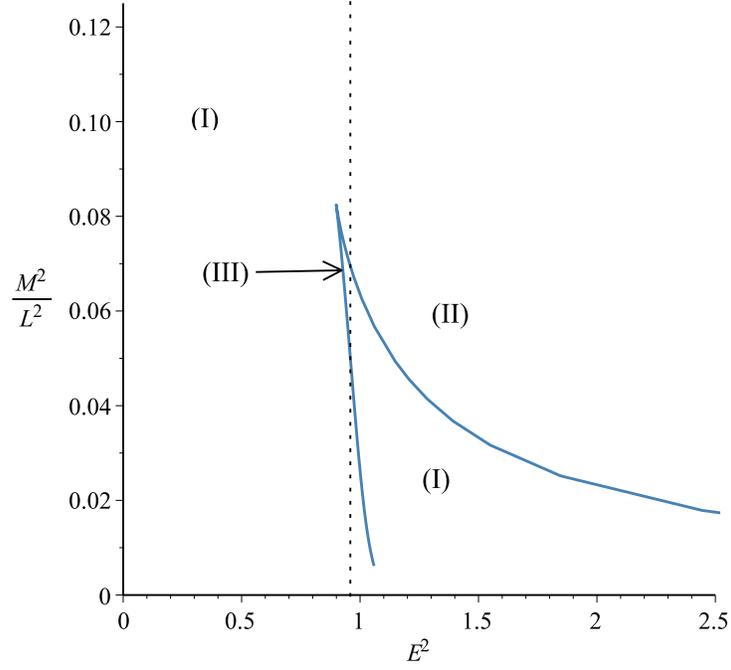}}
\centerline{\includegraphics[width=10cm]{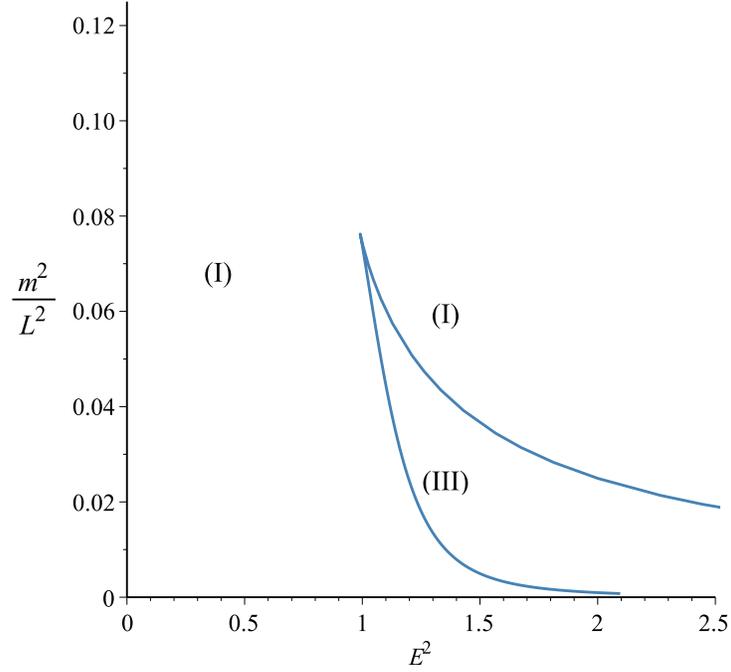}}
\caption{\label{region3}\small Regions of different types of geodesic motion for test particles ($ \epsilon=1 $). The numbers of positive real zeros in these
 regions are: (I)=2, (II)=0, (III)=3. For the effective potentials see Fig.~(\ref{potentials})}
\end{figure}

\begin{figure}[ht]
\centerline{\includegraphics[width=11cm]{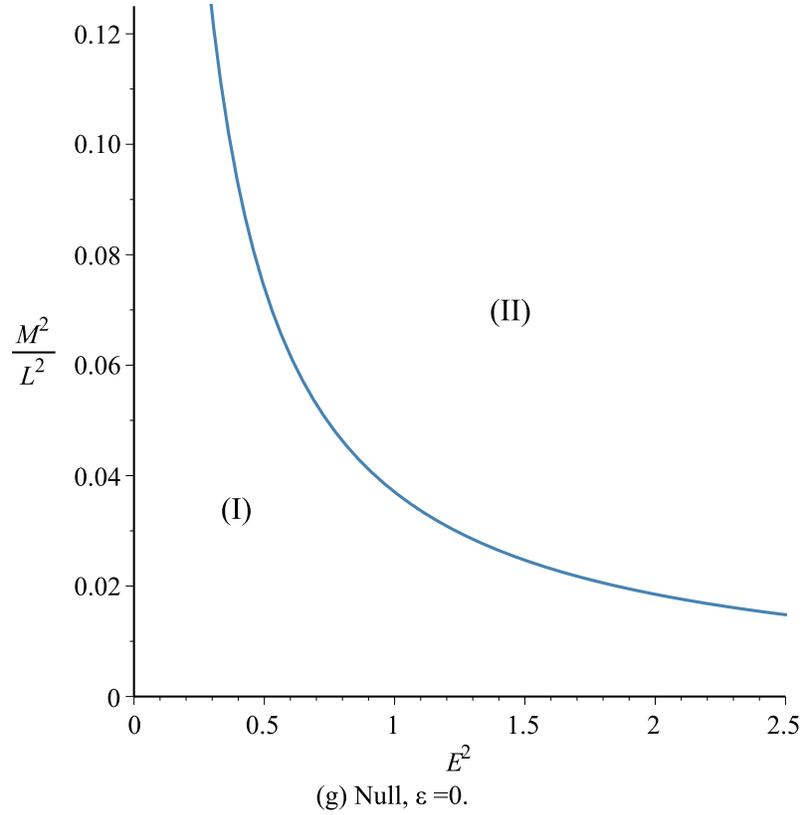}}
\caption{\label{region4}\small Regions of different types of geodesic motion for light ($ \epsilon=0 $). The numbers of positive real zeros in these
regions are: (I)=2, (II)=0. For the effective potentials see Fig.~(\ref{potentials})}
\end{figure}

If for a given set of parameters
$ \tilde{\Lambda}, \tilde{\beta}, \epsilon, E^{2} $, and $ \mathcal{L} $
the polynomial $ R^{*}(\tilde{r}) $ has $n$ positive
and real zeros, then for varying $ E^{2} $ and $L$
this number can only change if two zeros merge to one.
Solving $ R^{*}(\tilde{r})=0, \frac{d R^{*}(\tilde{r}) }{d\tilde{r}}=0  $ for $ E^{2} $  and $ \mathcal{L} $, for $ \epsilon=1 $ yields
\begin{equation}\label{LE^2T}
E^{2}=\frac{2(\tilde{\Lambda}\tilde{r}^{3}-\tilde{\beta}\tilde{r}^{2}-\tilde{r}+2)^{2}}{\tilde{r}(\tilde{\beta}\tilde{r}^{2}+2\tilde{r}-6)},\qquad\quad
\mathcal{L}=\frac{-(\tilde{\beta}\tilde{r}^{2}+2\tilde{r}-6)}{\tilde{r}^{2}(2\tilde{\Lambda}\tilde{r}^{3}-\tilde{\beta}\tilde{r}^{2}-2)}.
\end{equation}
and for $ \epsilon=0 $
\begin{equation}\label{LE^2N}
\mathcal{L}=\frac{1}{E^{2}}(\frac{4\tilde{\beta}^{3}}{(-1+\sqrt{1+6\tilde{\beta}})^{3}}-
\frac{\tilde{\beta}^{2}}{(-1+\sqrt{1+6\tilde{\beta}})^{2}}-\tilde{\Lambda}).
\end{equation}
In the limit $ \tilde{\beta}\rightarrow0 $ Eqs.(\ref{LE^2T}),(\ref{LE^2N}) reduce to the Schwarzschild-de Sitter expressions \cite{Hackmann:2008zz} $E^{2}=\frac{(\tilde{\Lambda}\tilde{r}^{3}-\tilde{r}+2)^{2}}{\tilde{r}(\tilde{r}-3)}, \mathcal{L}=\frac{-(\tilde{r}-3)}{\tilde{r}^{2}(\tilde{\Lambda}\tilde{r}^{3}-1)} $ and for $ \epsilon=0 $,
$\mathcal{L}=\frac{1}{E^{2}}(\frac{1}{27}-\tilde{\Lambda}) $
and in the  limit $ \tilde{\beta}\rightarrow0 $ and $\tilde{ \Lambda}\rightarrow0 $ reduce to the Schwarzschild expressions
$ E^{2}=\frac{(\tilde{r}-2)^{2}}{\tilde{r}(\tilde{r}-3)}, \mathcal{L}=\frac{(\tilde{r}-3)}{\tilde{r}^{2}} $ and for $ \epsilon=0 $, $\mathcal{L}=\frac{1}{27E^{2}} $ .
In Figs.~(\ref{region1},\ref{region2},\ref{region3} and \ref{region4})
the results of this analysis are shown for both test particles ($ \epsilon=1 $)
and light ($ \epsilon=0 $) for a small positive cosmological constant
$ \tilde{\Lambda} $ and $ \tilde{\beta} $ .

\clearpage

Three different regions of geodesic motion can be identified:
\begin{enumerate}
\item $R^{*}(\tilde{r}) $ has 2 positive real zeros $ r_{1}<r_{2} $ with $ R(\tilde{r})\geq 0 $ for  $ 0 \leq \tilde{r} \leq {r_{1}} $ and $ r_{2} \leq\tilde{r} $. Possible orbit types: flyby and terminating bound orbits.

\item $R^{*}(\tilde{r}) $ has 0 positive real zeros and $ R(\tilde{r})\geq 0 $ for positive $ \tilde{r} $. Possible orbit types: terminating escape orbits.

\item $R^{*}(\tilde{r}) $ has 4 positive real zeros $ r_{i}<r_{i+1} $ with $ R(\tilde{r})\geq 0 $ for  $ 0 \leq \tilde{r} \leq {r_{1}} $,
$ {r_{2}} \leq \tilde{r} \leq {r_{3}} $ and $ r_{4} \leq\tilde{r} $.
Possible orbit types: flyby, bound, and terminating bound orbits.

\end{enumerate}

\begin{figure}[ht]
\centerline{\includegraphics[width=7.25cm]{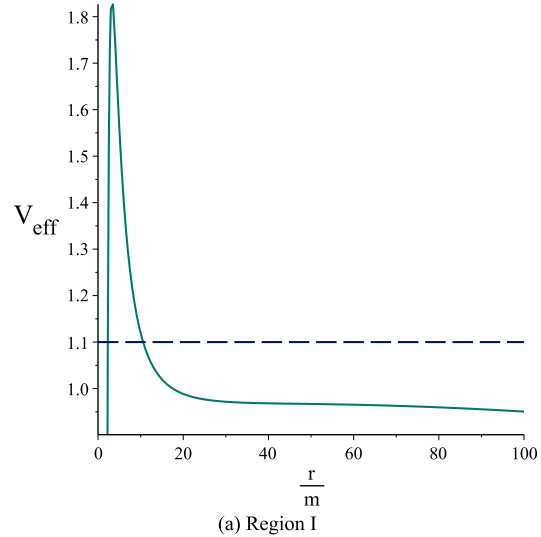}}
\centerline{\includegraphics[width=7.25cm]{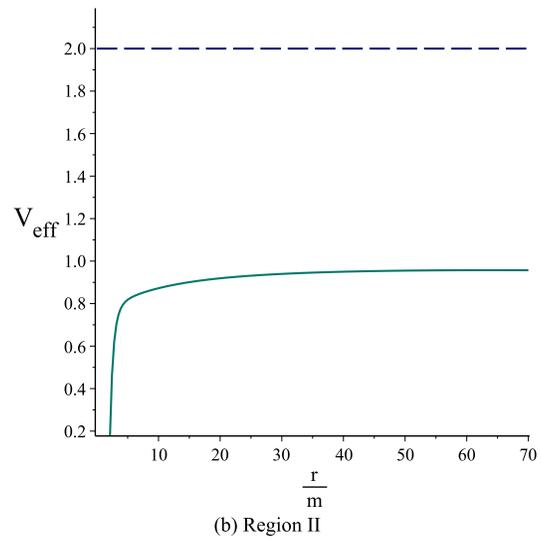}}
\centerline{\includegraphics[width=7.25cm]{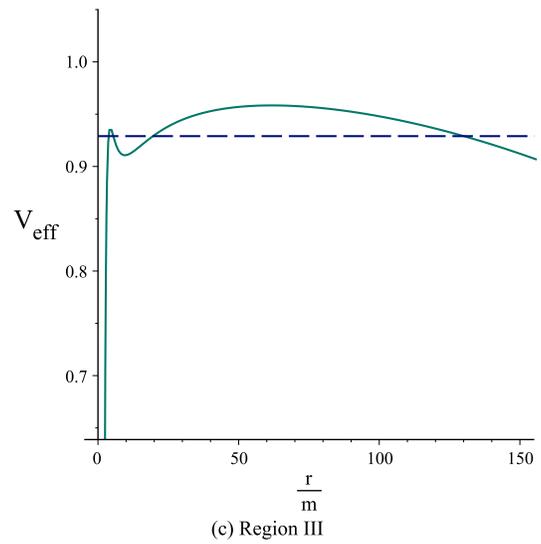}}
\caption{\label{potentials}\small Effective potentials for different regions of geodesic motion. The horizontal grey dashed line represents the squared energy parameter $ E^{2} $.}
\end{figure}


\begin{table}[ht]
\caption{\label{table}\small Types of orbits of light and particles in the spacetime of a black hole derived from the $f(R)$ modified gravity. The second column gives the number of positive zeros of the polynomial $ R^{*}(\tilde{r})$.
In the third column, the thick lines represent the range of orbits,
and the turning points are shown by thick dots.
The small vertical line denotes $ \tilde{r} $ = 0.}
\end{table}
\begin{center}

\vspace{3pt} \noindent
\begin{tabular}{|p{102pt}|p{102pt}|p{102pt}|p{102pt}|}
\hline
\parbox{102pt}{\centering
Region
} & \parbox{102pt}{\centering
Pos.Zeros
} & \parbox{102pt}{\centering
Range of $\tilde{r}$
} & \parbox{102pt}{\centering
Taype of Orbits
} \\
\hline
\parbox{102pt}{\centering
I
} & \parbox{102pt}{\centering
2
} & \parbox{102pt}{\centering
\begin{minipage}{.2\textwidth} \includegraphics[width=\linewidth, height=4mm]{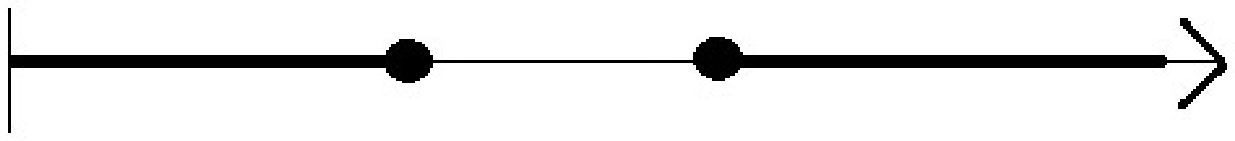} \end{minipage}
} & \parbox{102pt}{\centering
(TBO) , (FO)
} \\
\hline
\parbox{102pt}{\centering
II
} & \parbox{102pt}{\centering
0
} & \parbox{102pt}{\centering
\begin{minipage}{.2\textwidth}\includegraphics[width=\linewidth, height=4mm]{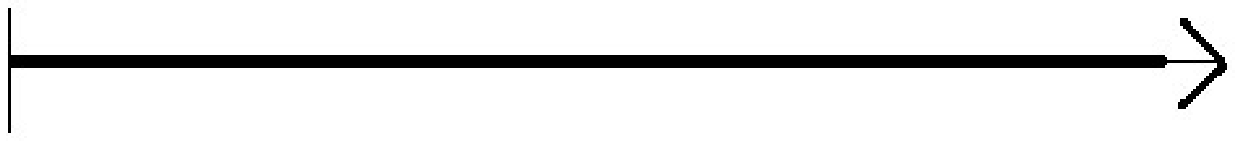} \end{minipage}
} & \parbox{102pt}{\centering
(TEO)
} \\
\hline
\parbox{102pt}{\centering
III
} & \parbox{102pt}{\centering
4
} & \parbox{102pt}{\centering
\begin{minipage}{.2\textwidth} \includegraphics[width=\linewidth, height=4mm]{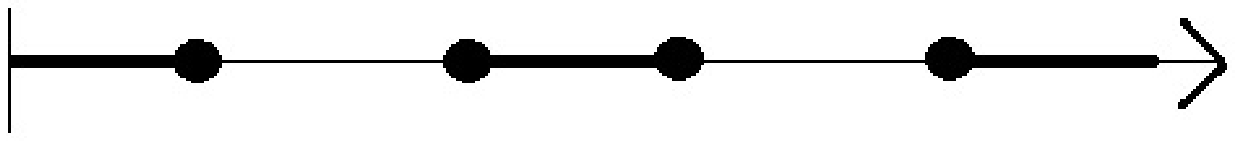} \end{minipage}
} & \parbox{102pt}{\centering
(TBO) , (BO) , (FO)
} \\
\hline
\end{tabular}
\vspace{2pt}

\end{center}

For light rays only regions (I) and (II) appear. Sample effective potentials for each
of the regions (I), (II), and (III) are shown in Fig.~(\ref{potentials}).
A summary of possible orbit types can be found in Tab.~(\ref{table}).
For light as well as test particles exceptional orbits
appear at the boundaries of the regions (I) to (III),
corresponding to multiple zeros of $R^{*}(\tilde{r})$.
In the case of multiple zeros the boundary is described by $E^{2}$
and $\mathcal{L}$ given by Eq.~(\ref{LE^2T}) or (\ref{LE^2N}), respectively,
and the corresponding test particle or light ray moves on a circular orbit,
which may be stable or unstable.
If $ R^{*}(\tilde{r}) $ has a maximum, then the circular orbit is stable
and else unstable. For test particles, the substitution of Eq.~(\ref{LE^2T})
in $ \frac{d^{2}R^{*}(\tilde{r})}{d\tilde{r}^{2}} $ yields
\begin{equation}\label{d^2R/dr^2}
\frac{d^{2}R^{*}(\tilde{r})}{d\tilde{r}^{2}}=\frac{-2(3\tilde{\Lambda}\tilde{\beta}\tilde{r}^{5}+(8\tilde{\Lambda}-\tilde{\beta}^{2})\tilde{r}^{4}-30\tilde{\Lambda}\tilde{r}^{3}+12\tilde{\beta}\tilde{r}^{2}-2\tilde{r}+12)}{\tilde{r}^{2}(2\tilde{\Lambda}\tilde{r}^{3}-\tilde{\beta}\tilde{r}^{2}-2)}.
\end{equation}

In the limit $ \tilde{\beta}=0 $ Eq.~(\ref{d^2R/dr^2}) reduces to the Schwarzschild-de Sitter expression \cite{Hackmann:2008zz}
$ \frac{d^{2}R^{*}(\tilde{r})}{d\tilde{r}^{2}}=\frac{-2(4\tilde{\Lambda}\tilde{r}^{4}-15\tilde{\Lambda}\tilde{r}^{3}-\tilde{r}+6)}{\tilde{r}^{2}(\tilde{\Lambda}\tilde{r}^{3}-1)} $
and in the limit $ \tilde{\beta}=0 $ and $ \tilde{\Lambda}=0 $
to the Schwarzschild expression
$ \frac{d^{2}R^{*}(\tilde{r})}{d\tilde{r}^{2}}=\frac{-2(\tilde{r}-6)}{\tilde{r}^{2}} $ .

The boundary of region (III) is defined by two corners, which correspond to the triple zeros of $ R^{*}(\tilde{r}) $. This means
that the region will vanish if triple zeros are not possible.
With an analysis similar to Eq.~(\ref{LE^2T})
it can be shown that triple zeros appear at
\begin{eqnarray}\label{triple}
E^{2}&=&\frac{8(\tilde{\beta}\tilde{r}^{2}+2\tilde{r}-6)^{3}}{\tilde{r}(3\tilde{\beta}\tilde{r}^{2}+8\tilde{r}-30)^{2}},\nonumber\\
\mathcal{L}&=&\frac{(3\tilde{\beta}\tilde{r}^{2}+8\tilde{r}-30)}{\tilde{r}^{2}(\tilde{\beta}\tilde{r}^{2}+6)},\nonumber\\  \tilde{\Lambda}&=&\frac{(\tilde{\beta}^{2}\tilde{r}^{4}+3\tilde{\beta}\tilde{r}^{3}-12\tilde{\beta}\tilde{r}^{2}+2\tilde{r}-12)}{\tilde{r}^{3}(3\tilde{\beta}\tilde{r}^{2}+8\tilde{r}-30)}.
\end{eqnarray}

In the limit $ \tilde{\beta}=0 $ Eqs.~(\ref{triple})
reduce to the Schwarzschild-de Sitter expressions
$E^{2}=\frac{16(\tilde{r}-3)^{2}}{\tilde{r}(4\tilde{r}-15)^{2}}, \mathcal{L}=\frac{4\tilde{r}-15}{3\tilde{r}^{2}}, \tilde{\Lambda}=\frac{(\tilde{r}-6)}{\tilde{r}^{3}(4\tilde{r}-15)} $.

Let us now compare the geodesic motion of test particles ($ \epsilon=1 $)
with the case $ \tilde{\beta}=0 $.
From Figs.~(\ref{region1},\ref{region2},\ref{region3} and \ref{region4})
it is obvious that some regions for small positive $ \tilde{\beta} $
are a little shifted compared to $ \tilde{\beta}=0 $.
But for the larger positive $ \tilde{\beta} $,
all regions for  $ \tilde{\beta}>0$  are shifted
compared to $\tilde{\beta}=0 $.
Thus, every pair $ (E^{2}, \mathcal{L})$,
which was located near a boundary for $\tilde{\beta}=0 $,
may switch to another region.
Regions (IV) and (III) absorbed the whole parameter space,
which was located in regions (II) and (I), respectively,
for $\tilde{\beta}=0 $, see Fig.~(\ref{region1},\ref{region2}: a,c)
and for cases with the positive cosmological constant,
regions (I) absorbed the whole parameter space,
which was located in regions (II) for $\tilde{\beta}=0 $.
For  $\tilde{\beta}=0 $, there are also terminating escape orbits which,
coming from infinity, are reflected at the potential barrier
induced by the larger positive  $\tilde{\beta} $.
The types of orbits for null geodesics, i.e., for light rays,
are not essentially influenced
by a nonvanishing $\tilde{\beta}$.
For every region, examples of timelike and null geodesics can be found in Figs.~(\ref{OT1},\ref{OT2},\ref{OT3},\ref{ON1},\ref{ON2},\ref{OC1},\ref{OC2}) .

\subsection{Analytical solution of geodesic equations}

In this section we present the analytical solution of the equations of motion.
We introduce a new variable $ u=\frac{1}{\tilde{r}} $,
and obtain from Eq.(\ref{dr/dphin}):
\begin{equation}\label{du/dphi}
(\frac{du}{d\varphi})^2=2u^{3}-u^{2}+(2\epsilon\mathcal{L}-\tilde{\beta})u+((E^{2}-\epsilon)\mathcal{L}+\tilde{\Lambda})\\
-\epsilon\tilde{\beta}\mathcal{L}\frac{1}{u}+\epsilon\tilde{\Lambda}\mathcal{L}\frac{1}{u^{2}} .
\end{equation}

\subsubsection{\textbf{Null geodesics}}

For $ \epsilon = 0 $ Eq.~(\ref{du/dphi}) is of elliptic type
\begin{equation}\label{du/dphiN}
(\frac{du}{d\varphi})^2=2u^{3}-u^{2}-\tilde{\beta}u+(E^{2}\mathcal{L}+\tilde{\Lambda})=
p_{3}(u)=\sum_{i=0}^{3}a_{i}u^{i}.
\end{equation}
A further substitution $ u=\frac{1}{a_{3}}(4y-\frac{a_{2}}{3})=2y+\frac{1}{6} $ transforms $ p_{3}(u) $ into the Weierstrass form so that Eq.~(\ref{du/dphiN}) turns into:
 \begin{equation}\label{Weierstrass function}
(\frac{dy}{d\varphi})^{2}=4y^{3}-g_{2}y-g_{3}=p_{3}(y),
\end{equation}
 where
 \begin{equation}
g_{2}=\frac{a_{2}^{2}}{12}-\frac{a_{1}a_{3}}{4}=\frac{1}{12}+\frac{\tilde{\beta}}{2} ,\quad  g_{3}=\frac{a_{1}a_{2}a_{3}}{48}-\frac{a_{0}a_{3}^{2}}{16}-\frac{a_{2}^{3}}{216}=
\frac{\tilde{\beta}}{24}-\frac{(E^{2}\mathcal{L}+\tilde{\Lambda})}{4}+\frac{1}{216} \quad
\end{equation}
are the Weierstrass invariants.
The differential equation (\ref{Weierstrass function}) is of elliptic type
and is solved by the Weierstrass function
\cite{Hackmann:2008zz,Abramowitz:1968,Whittaker:1973}.
\begin{equation}
y(\varphi)=\wp(\varphi-\varphi_{in};g_{2},g_{3}),
\end{equation}
where $\varphi_{in}=\varphi_{0}+\int_{y_{0}}^{\infty} \frac{dy}{\sqrt{4y^{3}-g_{2}y-g_{3}}}$
with $ y_{0}=\frac{a_{3}}{4\tilde {r_{0}}}+\frac{a_{2}}{12}=\frac{1}{2}(\frac{1}{\tilde {r_{0}}}-\frac{1}{6}) $, depending only on the initial values $ \varphi_{0} $ and $ \tilde{r_{0}} $.
Then the solution of (\ref{dr/dphin}) acquires the form
\begin{equation}
\tilde{r}(\varphi)=\frac{a_{3}}{4\wp(\varphi-\varphi_{in};g_{2},g_{3})-\frac{a_{2}}{3}}=\frac{1}{2\wp(\varphi-\varphi_{in};g_{2},g_{3})+\frac{1}{6}} .
\end{equation}
The corresponding light trajectories have been exhaustively
discussed in \cite{Hagihara:1931}.\\

\subsubsection{\textbf{Timelike geodesics}}
For $ \epsilon=1$ Eq.(\ref{du/dphi}) should be rewritten as
\begin{equation}\label{Timelike}
(u\frac{du}{d\varphi})^2=2u^{5}-u^{4}+(2\mathcal{L}-\tilde{\beta})u^{3}+((E^{2}-1)\mathcal{L}+\tilde{\Lambda})u^{2}
-\tilde{\beta}\mathcal{L}u+\tilde{\Lambda}\mathcal{L}=p_{5}(u)=\sum_{i=0}^{5}a_{i}u^{i}.
\end{equation}
A separation of variables in (\ref{Timelike}) yields
\begin{equation}\label{Phi}
\varphi - \varphi_{0}=\int_{u_{0}}^{u}\frac{udu}{\sqrt{p_{5}(u)}},
\end{equation}
where $ u_{0}=u(\varphi_{0}) $. In solving integral (\ref{Phi}) there are two major issues which have to be
addressed. First, the integrand is not well defined in the complex plane because of the two
branches of the square root. Second, the solution $ u(\varphi) $ should not depend on the integration
path \cite{Hackmann:2008zz}. If $ \gamma $ denotes some closed integration path and
\begin{equation}\label{Omega}
\omega=\oint_{\gamma}\frac{udu}{\sqrt{p_{5}(u)}},
\end{equation}
this means that
\begin{equation}\label{PhiOmega}
\varphi - \varphi_{0}-\omega=\int_{u_{0}}^{u}\frac{udu}{\sqrt{p_{5}(u)}}
\end{equation}
should be valid, too. Hence, the solution $ u(\varphi) $ of our problem has to fulfill
\begin{equation}\label{PO}
u(\varphi)=u(\varphi - \omega)
\end{equation}
for every $ \omega\neq{0} $ obtained from Eq.~(\ref{Omega}). These two issues can be solved if
we consider Eq. (\ref{Phi}) to be defined on the Riemann surface $ y^{2}=p_{5}(x) $ of genus $ g=2 $
and introduce a basis of canonical holomorphic and meromorphic differentials $ dz_{i} $ and $ dr_{i} $,
respectively,
\begin{equation}\label{dz}
dz_{1}=\frac{dx}{\sqrt{p_{5}(x)}}, \qquad\qquad\qquad dz_{2}=\frac{xdx}{\sqrt{p_{5}(x)}},
\end{equation}
\begin{equation}\label{dr}
dr_{1}=\frac{a_{3}x+2a_{4}x^{2}+3a_{5}x^{3}}{4\sqrt{p_{5}(x)}}dx, \qquad\qquad\qquad dr_{2}=\frac{x^{2}dx}{4\sqrt{p_{5}(x)}},
\end{equation}
and real $ 2\omega_{ij} $ , $ 2\eta_{ij} $ and imaginary $ 2\omega^{\prime}_{ij} $ , $ 2\eta^{\prime}_{ij} $ period matrices
\begin{equation}\label{2w}
2\omega_{ij}=\oint_{a_{j}} {dz_{i}}, \qquad\qquad\qquad 2\omega^{\prime}_{ij}=\oint_{b_{j}} {dz_{i}},
\end{equation}
\begin{equation}\label{2n}
2\eta_{ij}=\oint_{a_{j}} {dr_{i}}, \qquad\qquad\qquad 2\eta^{\prime}_{ij}=\oint_{b_{j}} {dr_{i}}.
\end{equation}
Now we can write the analytic solution of equation (\ref{Timelike}) as \cite{Hackmann:2008zz,Enolski:2010if}

\begin{equation}\label{ Kleinian}
u(\varphi)=-\frac{\sigma_{1}}{\sigma_{2}}(\varphi_{\sigma}),
\end{equation}
where $ \sigma_{i} $ is the $i$-th derivative of the Kleinian sigma function in two variables
\begin{equation}
\sigma(z)=Ce^{z^{t}kz} \theta[K_{\infty}](2\omega^{-1}z;\tau),
\end{equation}
which is given by the Riemann $\theta $-function with characteristic
$K_{\infty} $,
\begin{equation}
\theta(z;\tau)=\sum_{m\in \mathbb{Z}^{2}}e^{i{\pi}m^{t}({\tau}m+2z)}.
\end{equation}

A number of parameters
enter here: the symmetric Riemann matrix $\tau=(\omega^{-1}\omega^{\prime}) $, the period-matrix $(2\omega, 2\omega^{\prime}) $, the period-matrix
of the second kind $ (2\eta, 2\eta^{\prime}) $, the matrix $ k=\eta(2\omega)^{-1} $, and the vector of Riemann
constants with base point at infinity $ 2K_{\infty} = (0, 1)^{t} + (1, 1)^{t}\tau$. The constant $C$ can be
given explicitly, see e.g.\cite{Buchstaber:1997}, but does not matter here.
In Eq.~(\ref{ Kleinian}) the argument $ \varphi_{\sigma} $  is an
element of the one-dimensional sigma divisor: $ \varphi_{\sigma} = (f(\varphi -\varphi_{in}),\varphi -\varphi_{in})^{t} $ where
$\varphi_{in}=\varphi_{0}+\int_{u_{0}}^{\infty} \frac{udu}{\sqrt{p_{5}(u)}}$
with $ u_{0}=\frac{1}{\tilde{r_{0}}}$ depends only on the initial values,
 and the function $f$ is given by the condition $ \sigma(\varphi_{\sigma})=0 $. For more details on the construction of
such solutions see e.g.\cite{Enolski:2010if,Buchstaber:1997}.
With Eq.(\ref{ Kleinian}) the solution for $ \tilde{r} $ is given by
\begin{equation}
\tilde{r}=-\frac{\sigma_{2}}{\sigma_{1}}(\varphi_{\sigma}).
\end{equation}
In this equation, $ \sigma_{1} $ and $ \sigma_{2} $ depend on the
set of parameters $ \varphi_{\sigma},\omega, \eta, \tau $ and also
on $ p_{5}(u) $ according to Eqs.~($ \ref{Phi}-\ref{2n} $), which
contains the $\beta$-dependence of the modified gravity. The
solution of $ \tilde{r} $ is the analytic solution of the equation
of motion of a test particle in the spacetime of a black hole
derived in $f(R)$ modified gravity. This solution is valid in all
regions of this spacetime.
\\

\subsubsection{\textbf{The orbits}}

\begin{figure}[ht]
\centerline{\includegraphics[width=10cm]{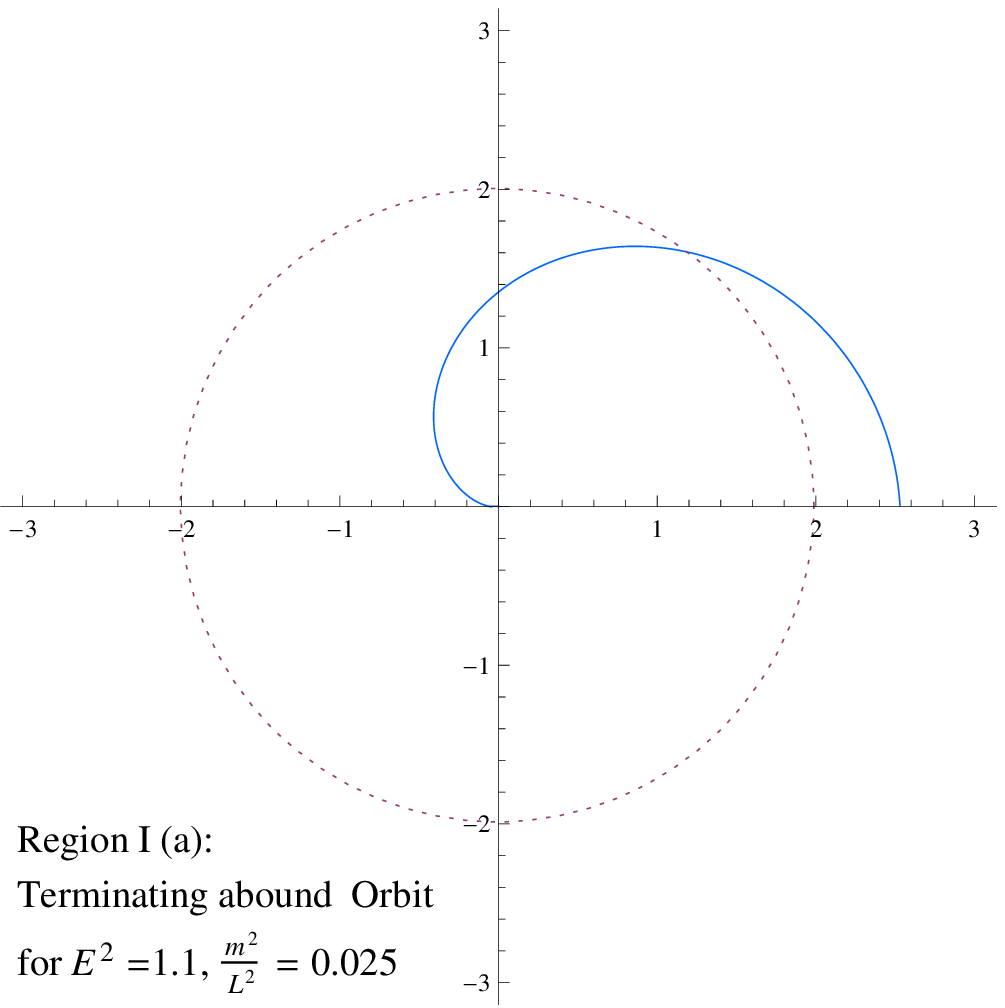}}
\centerline{\includegraphics[width=10cm]{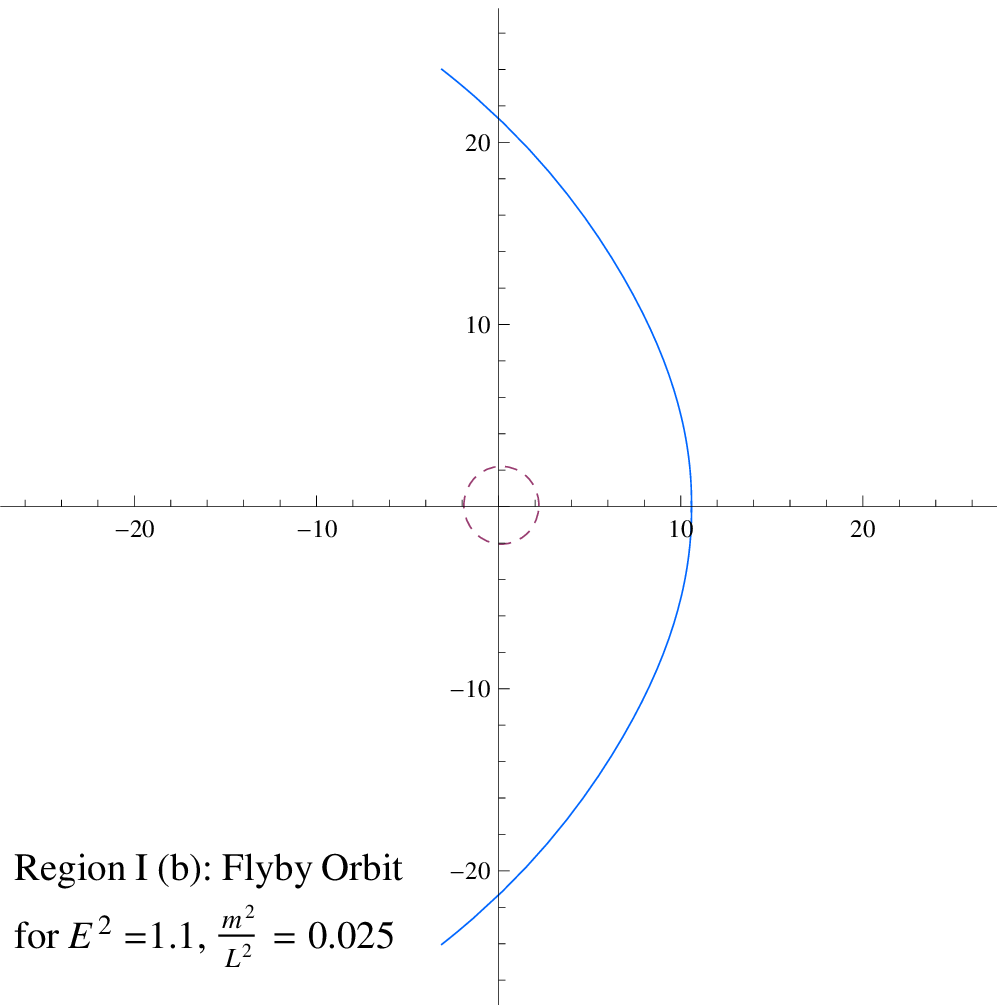}}
\caption{\label{OT1}\small  Timelike geodesics for the different regions of orbit types $ (\tilde{\Lambda}=\frac{1}{3}(10^{-5}), \tilde{\beta}=10^{-3} ) $.  Circles always indicate the Schwarzschild radius.}
\end{figure}

\begin{figure}[ht]
\centerline{\includegraphics[width=10cm]{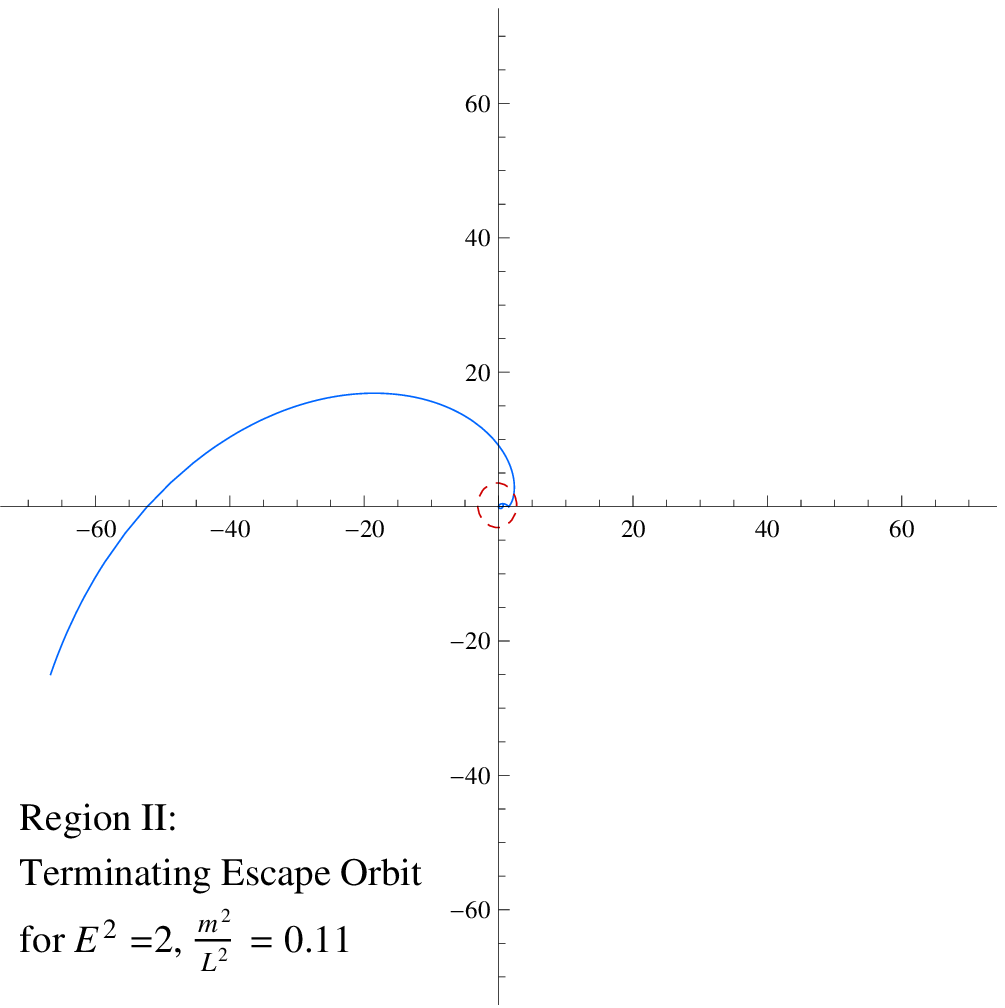}}
\centerline{\includegraphics[width=10cm]{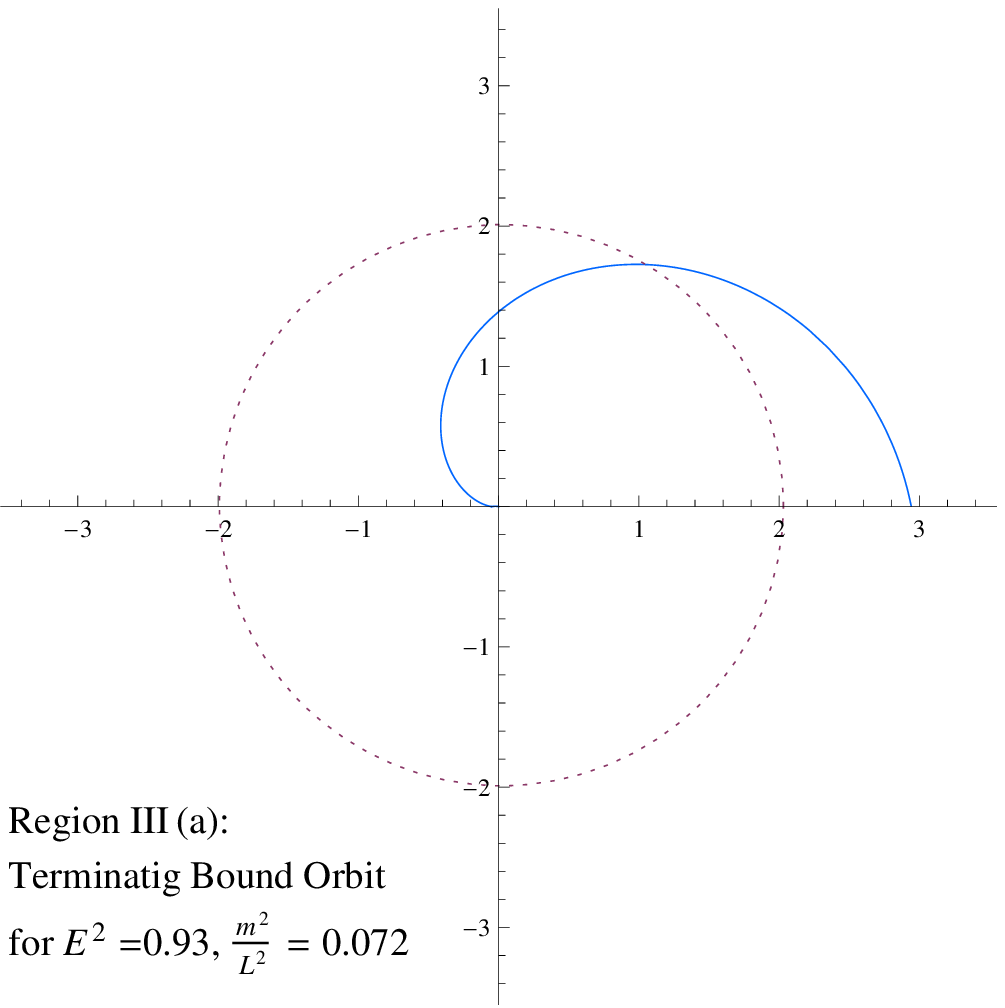}}
\caption{\label{OT2}\small  Timelike geodesics for the different regions of orbit types $ (\tilde{\Lambda}=\frac{1}{3}(10^{-5}), \tilde{\beta}=10^{-3} ) $.  Circles always indicate the Schwarzschild radius.}
\end{figure}

\begin{figure}[ht]
\centerline{\includegraphics[width=10cm]{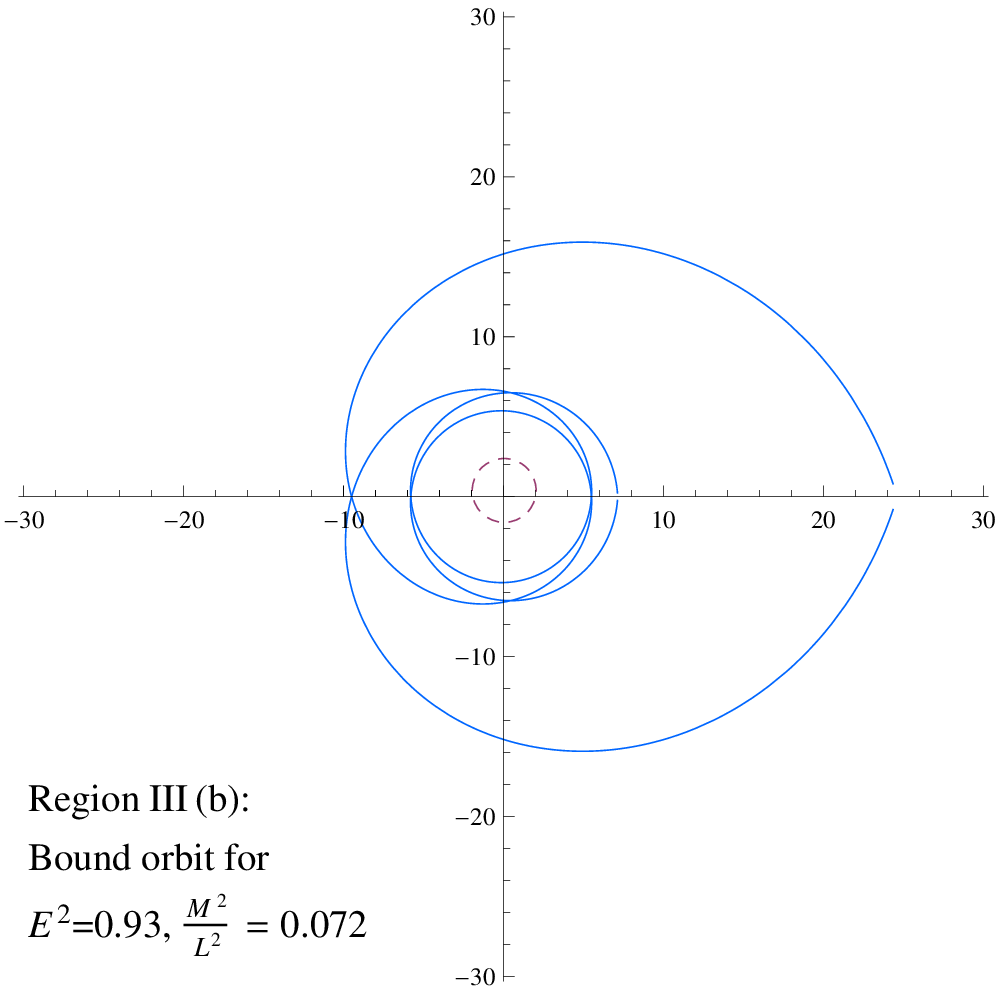}}
\centerline{\includegraphics[width=10cm]{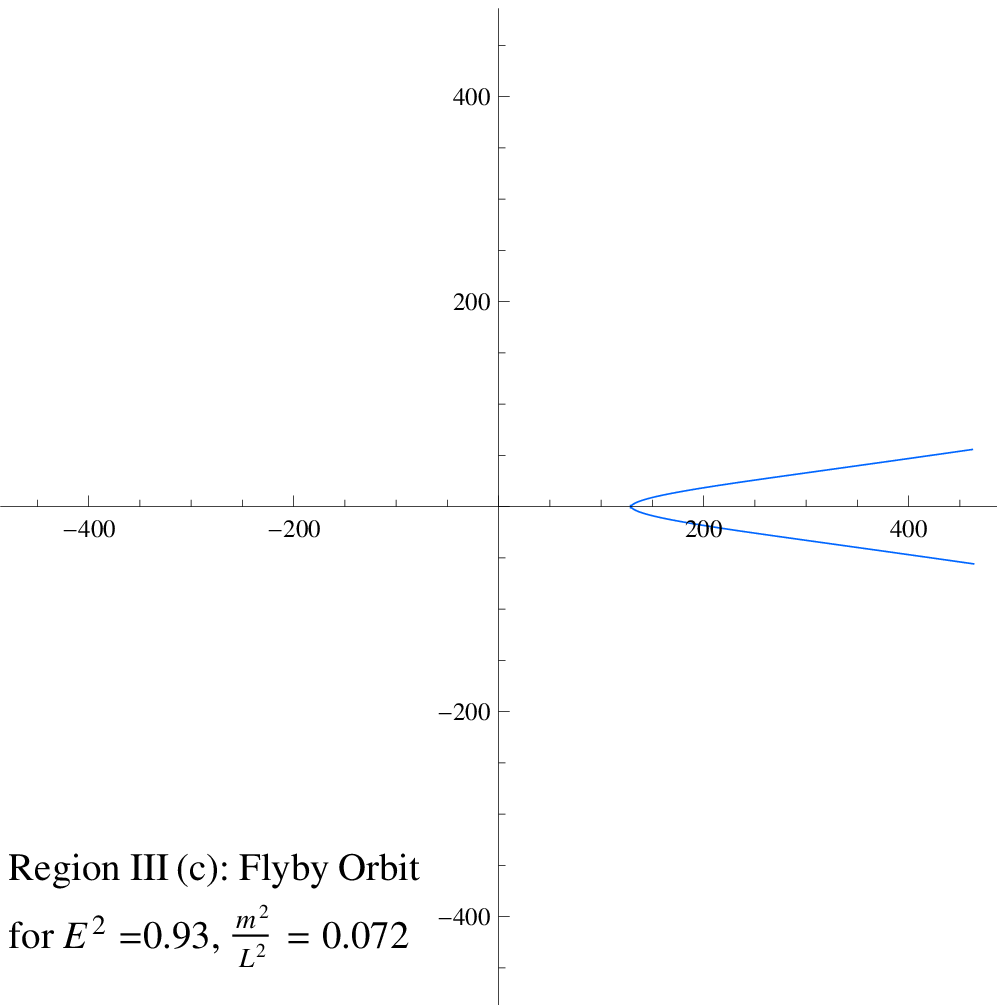}}
\caption{\label{OT3}\small  Timelike geodesics for the different regions of orbit types $ (\tilde{\Lambda}=\frac{1}{3}(10^{-5}), \tilde{\beta}=10^{-3} ) $.  Circles always indicate the Schwarzschild radius.}
\end{figure}

\begin{figure}[ht]
\centerline{\includegraphics[width=10cm]{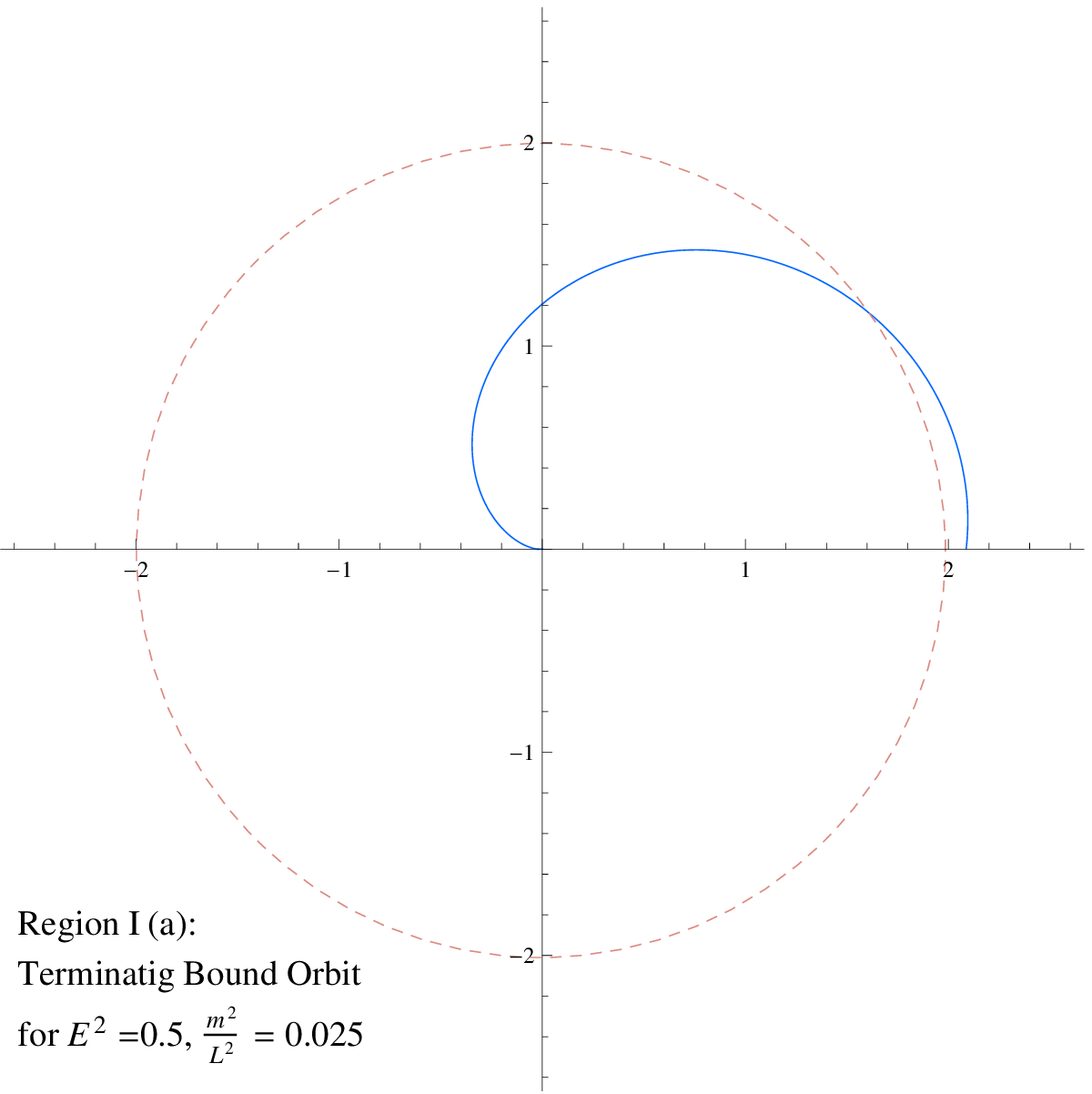}}
\centerline{\includegraphics[width=10cm]{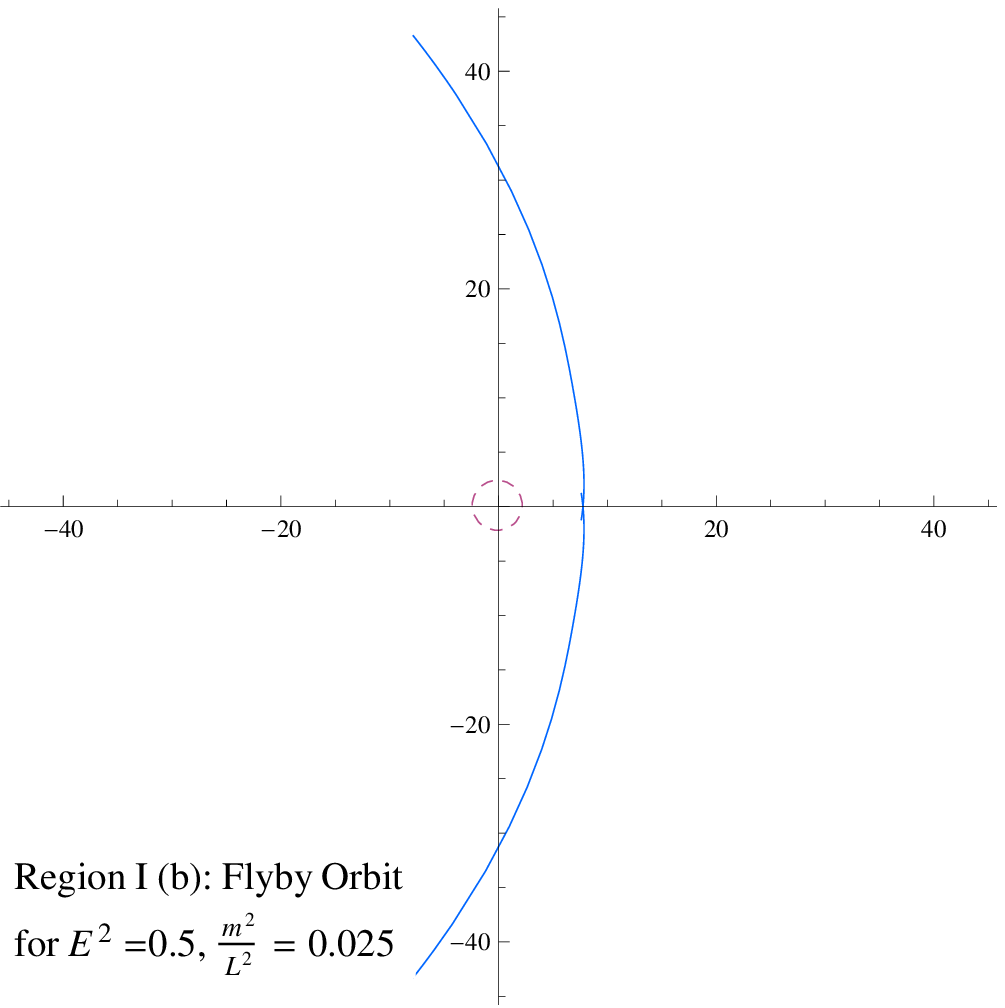}}
\caption{\label{ON1}\small Null geodesics for the different regions of orbit types $ (\tilde{\Lambda}=\frac{1}{3}(10^{-5}), \tilde{\beta}=10^{-3} ) $.}
\end{figure}

\begin{figure}[ht]
\centerline{\includegraphics[width=11cm]{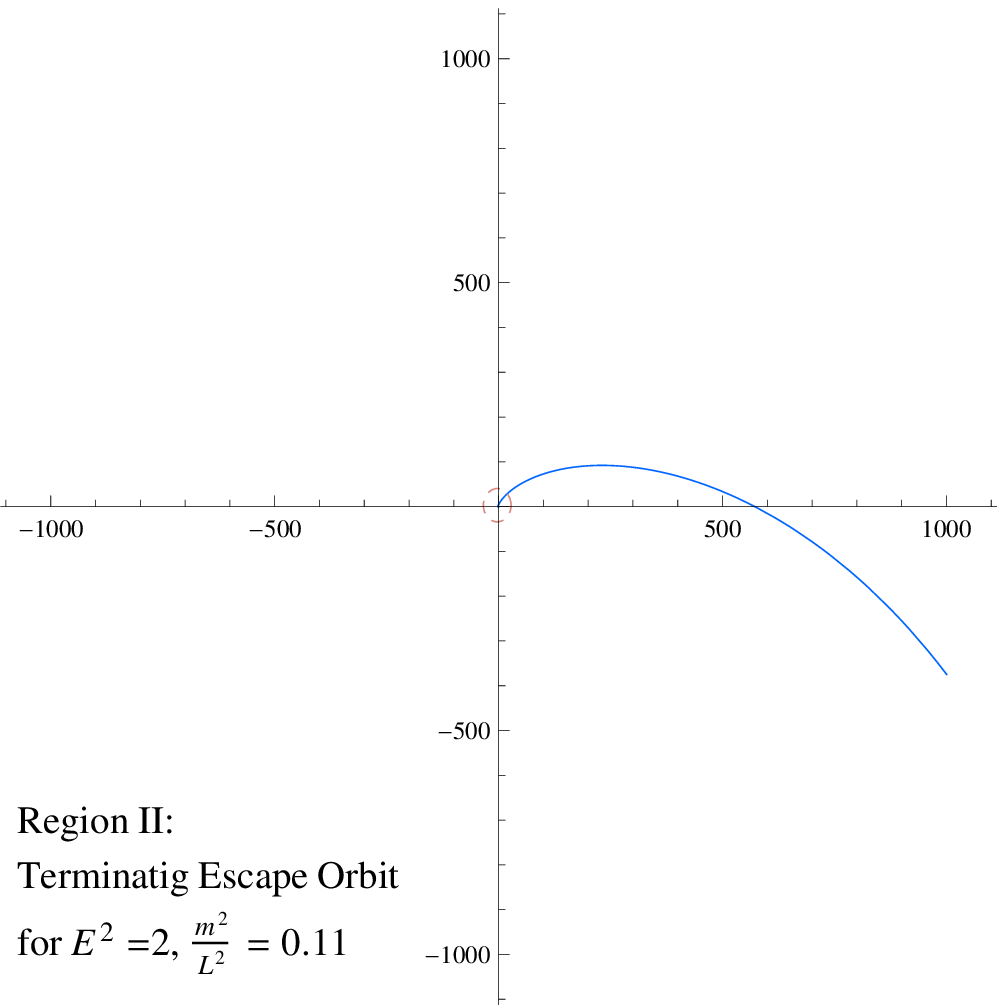}}
\caption{\label{ON2}\small Null geodesics for the different regions of orbit types $ (\tilde{\Lambda}=\frac{1}{3}(10^{-5}), \tilde{\beta}=10^{-3} ) $.}
\end{figure}

\begin{figure}[ht]
\centerline{\includegraphics[width=10cm]{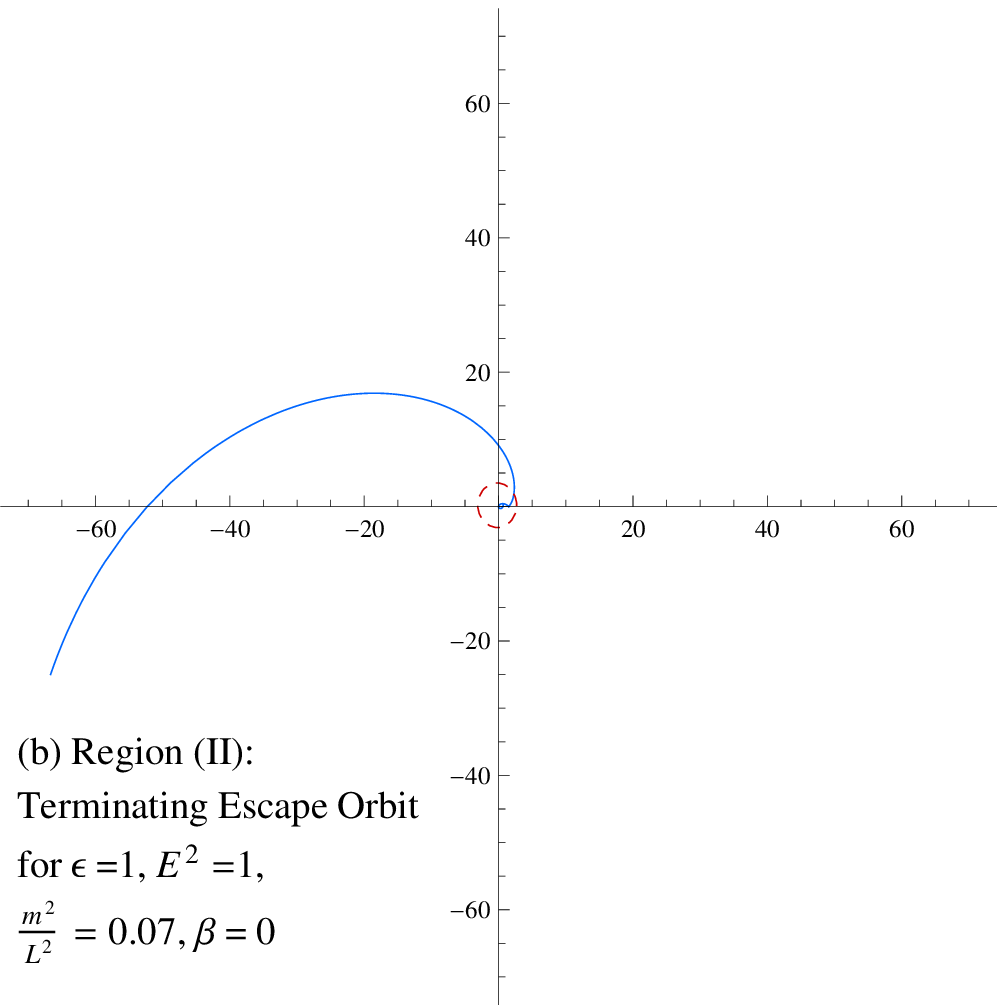}}
\centerline{\includegraphics[width=10cm]{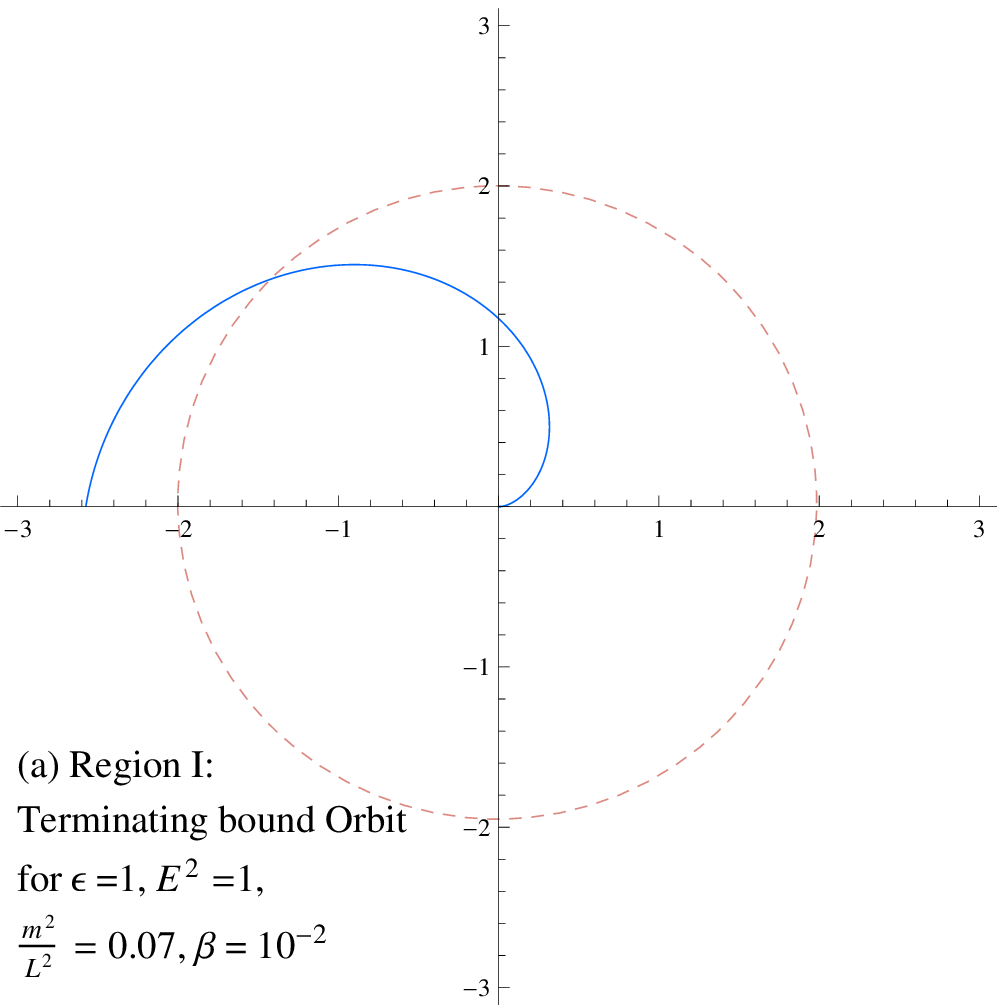}}
\caption{\label{OC1}\small Comparison between orbits in the cases of
nonvanishing $\tilde{\beta}$ and $ \tilde{\beta}=0 $.}
\end{figure}

\begin{figure}[ht]
\centerline{\includegraphics[width=10cm]{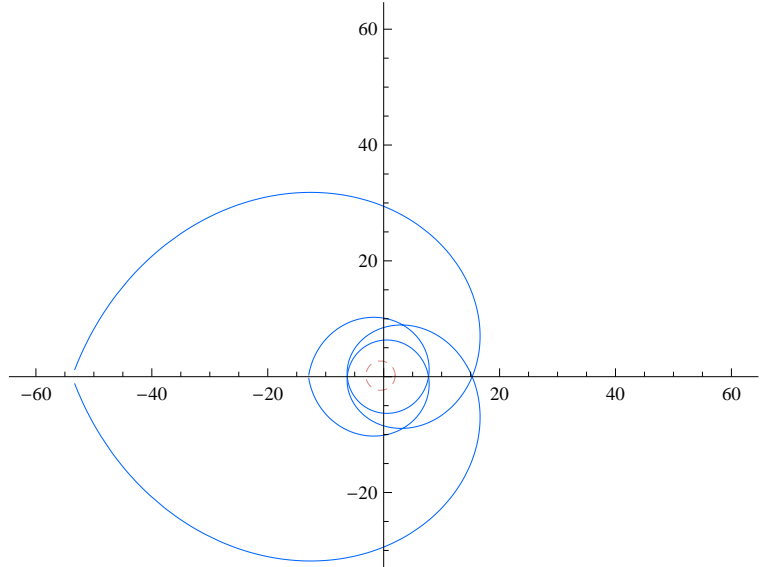}}
\centerline{\includegraphics[width=10cm]{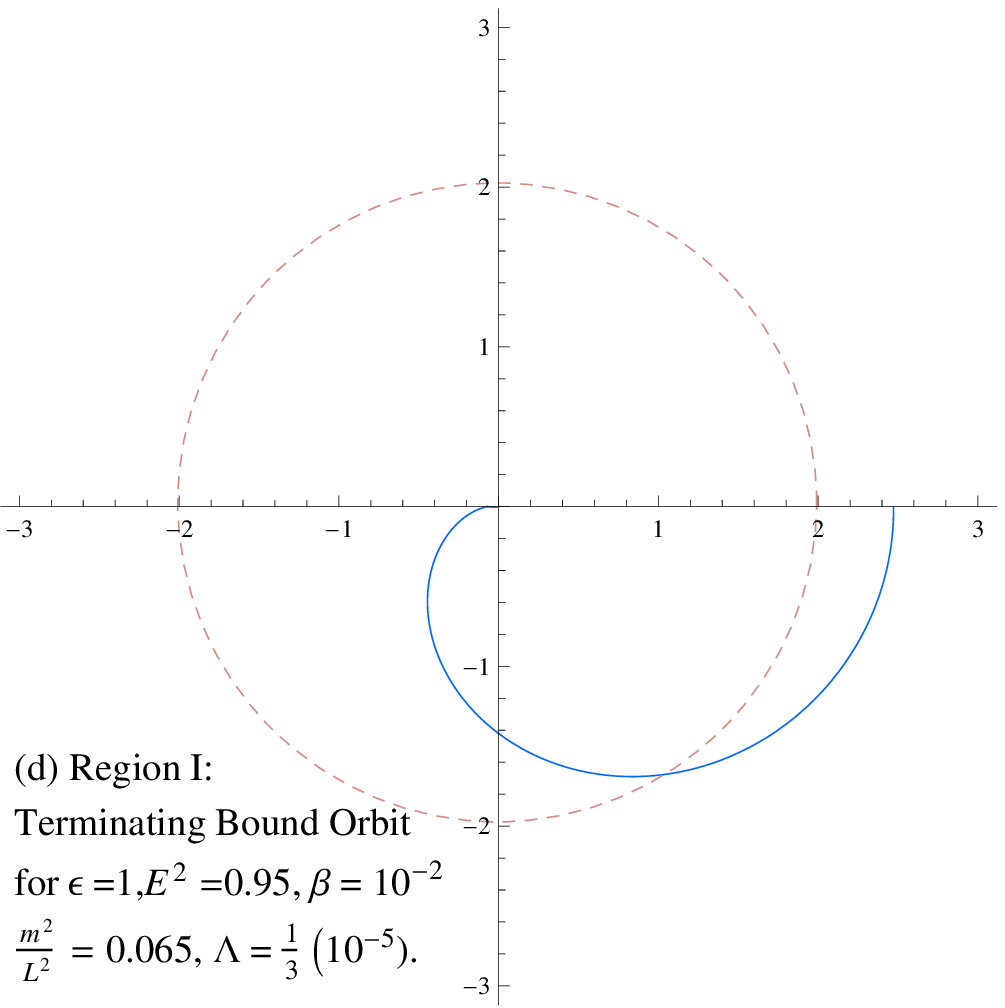}}
\caption{\label{OC2}\small Comparison between orbits in the cases of
nonvanishing $\tilde{\beta}$ and $ \tilde{\beta}=0 $.}
\end{figure}

In Figs.~(\ref{OT1},\ref{OT2},\ref{OT3},\ref{ON1},\ref{ON2},\ref{OC1},\ref{OC2}) some of the possible orbits are plotted.
For all orbits in each region of the figure the parameters  $E$
and $\mathcal{L}$ are the same.
The value of the cosmological constant is chosen as
$ \tilde{\Lambda}=\frac{1}{3}(10^{-5}) $ in all plots.
For massive test particles:
In Fig.~(\ref{OT1}) the parameters $ E^{2}=1.1$ and $\mathcal{L}=0.025 $
belong to the (I) region in Fig.~(\ref{region3})(e)
and Fig.~(\ref{potentials})(a).
For a positive cosmological constant 
this corresponds to a flyby orbit and a bound terminating orbit,
ending in the singularity.
The corresponding orbits are shown in Figs.~(\ref{OT1})(a),(\ref{OT1})(b).
The next parameter choice is $ E^{2}=2$ and $\mathcal{L}=0.11 $.
These parameters lie in the (II) region of Fig.~(\ref{region3})(e)
and Fig.~(\ref{potentials})(b) corresponding to a terminating escape orbit
ending in the singularity, Fig.~(\ref{OT2}).
Our third choice of parameters is $ E^{2}=0.93$ and $\mathcal{L}=0.072 $.
These parameters are in the (III) region in Fig.~(\ref{region3})(e)
and Fig.~(\ref{potentials})(c)
with four zeros, indicating a terminating bound orbit, a bound orbit
and a flyby orbit, see Figs.~(\ref{OT2})(a),(\ref{OT3})(b) and (\ref{OT3})(c).

The dependence of the perihelion shift of a bound orbit of a massive
test particle is shown in Fig.~(\ref{OT3})(b).
The periastron advance for such a bound orbit is given by the difference
of the $2\pi$-periodicity of the angle $\varphi$ and the periodicity
of the solution $r(\varphi)$. For more details on the construction
of such solution see e.g.~\cite{Hackmann:2008zz,Hartmann:2010rr}.
The orbit in Fig.~(\ref{OT3})(c) is a reflection at the $ \Lambda $ barrier.

For light rays: In Figs.~(\ref{ON1},\ref{ON2}) some of the possible orbits
are plotted. The parameters $ E^{2}=0.5 $ and $\mathcal{L}=0.025 $
belong to the (I) region in Fig.~(\ref{region4})(b)
and Fig.~(\ref{potentials})(a), corresponding to a flyby orbit
and a bound terminating orbit ending in the singularity.
The corresponding orbits are shown in Figs.~(\ref{ON1})(a),(\ref{ON1})(b).
The next parameter choice is $ E^{2}=2$ and $\mathcal{L}=0.11 $.
These parameters lie in the (II) area of Fig.~(\ref{region4})(b)
and Fig.~(\ref{potentials})(b), corresponding to a terminating escape orbit
ending in the singularity, Fig.~(\ref{ON2}).
The deflection of light (massless test particle, i.e. $ \epsilon=0 $)
is shown in Fig.~(\ref{ON1})(b).
For more details on the deflection of light solution
see e.g.~\cite{Gibbons:2011rh,Hartmann:2010rr}.
Some of the types of orbits for timelike geodesics which are
essentially influenced by a nonvanishing $\tilde{\beta}$
are presented in Figs.~(\ref{OC1},\ref{OC2}).

\clearpage

\section{CONCLUSIONS}\label{conclusions}

In this paper we considered the motion of massive and massless test particles
in a black hole metric of $f(R)$ gravity, presented in \cite{Saffari:2007zt}.
After reviewing the spacetime and the corresponding equations of motion
we classified the complete set of orbit types for massive and massless
test particles moving on geodesics.
In particular, we analyzed conditions for circular orbits.
The geodesic equations can be solved in terms of
Weierstrass elliptic functions and derivatives of Kleinian sigma functions.

We also  considered all possible types of orbits.
Using effective potential techniques and parametric diagrams,
the possible types of orbits were derived.
For null geodesics FO, TBO and TEO are possible,
while for timelike geodesics BO, FO, TBO and TEO are possible.
We also performed a comparison between orbits in the cases
of nonvanishing $\tilde{\beta}$ and $ \tilde{\beta}=0 $.

The results obtained in this paper present a useful
tool to calculate the exact orbits and their properties,
including observables like the periastron shift of bound orbits,
the light deflection of flyby orbits, the
deflection angle and the Lense-Thirring effect.
For the calculation of the observables analogous formulas to those given
in ~\cite{Hackmann:2008zz, Hartmann:2010rr, Gibbons:2011rh} can be used.
It would be interesting to extend this work to a charged and rotating
version of the black hole spacetime, which has still to be derived.

\begin{acknowledgments}
J.K.~and C.L.~would like to acknowledge support by the DFG
Research Training Group {\sl Model of Gravity}.
\end{acknowledgments}


\begin{thebibliography}{99}
\bibitem{Riess:1998cb}
  A.~G.~Riess {\it et al.}  [Supernova Search Team Collaboration],
  Astron.\ J.\  {\bf 116}, 1009 (1998)
  [astro-ph/9805201].
  S.~Perlmutter {\it et al.}  [Supernova Cosmology Project Collaboration],
  Astrophys.\ J.\  {\bf 517}, 565 (1999)
  [astro-ph/9812133].
  J.~L.~Tonry {\it et al.}  [Supernova Search Team Collaboration],
  Astrophys.\ J.\  {\bf 594}, 1 (2003)
  [astro-ph/0305008].
  C.~L.~Bennett {\it et al.}  [WMAP Collaboration],
  Astrophys.\ J.\ Suppl.\  {\bf 148}, 1 (2003)
  [astro-ph/0302207].
  G.~Hinshaw {\it et al.}  [WMAP Collaboration],
  Astrophys.\ J.\ Suppl.\  {\bf 170}, 288 (2007)
  [astro-ph/0603451].
  G.~Hinshaw {\it et al.}  [WMAP Collaboration],
  Astrophys.\ J.\ Suppl.\  {\bf 170}, 288 (2007)
  [astro-ph/0603451].
\bibitem{Sahni:1999gb}
  V.~Sahni and A.~A.~Starobinsky,
  Int.\ J.\ Mod.\ Phys.\ D {\bf 9}, 373 (2000)
  [astro-ph/9904398].
\bibitem{Carroll:2000fy}
S.~M.~Carroll,
  Living Rev.\ Rel.\  {\bf 4}, 1 (2001)
  [astro-ph/0004075].
  P.~J.~E.~Peebles and B.~Ratra,
  Rev.\ Mod.\ Phys.\  {\bf 75}, 559 (2003)
  [astro-ph/0207347].
  T.~Padmanabhan,
  Phys.\ Rept.\  {\bf 380}, 235 (2003)
  [hep-th/0212290].
  E.~J.~Copeland, M.~Sami and S.~Tsujikawa,
  Int.\ J.\ Mod.\ Phys.\ D {\bf 15}, 1753 (2006)
  [hep-th/0603057].
\bibitem{Kagramanova:2006ax}
V.~Kagramanova, J.~Kunz and C.~L\"ammerzahl,
  Phys.\ Lett.\ B {\bf 634}, 465 (2006).
\bibitem{Hackmann:2008zz}
  E.~Hackmann and C.~Lammerzahl,
  Phys.\ Rev.\ D {\bf 78}, 024035 (2008).
\bibitem{Carroll:2003wy}
  S.~M.~Carroll, V.~Duvvuri, M.~Trodden and M.~S.~Turner,
  Phys.\ Rev.\ D {\bf 70}, 043528 (2004)
  [astro-ph/0306438].
  S.~Nojiri and S.~D.~Odintsov,
  Phys.\ Rev.\ D {\bf 68}, 123512 (2003)
  [hep-th/0307288].
  S.~Nojiri and S.~D.~Odintsov,
  Gen.\ Rel.\ Grav.\  {\bf 36}, 1765 (2004)
  [hep-th/0308176].
  S.~Nojiri and S.~D.~Odintsov,
  Mod.\ Phys.\ Lett.\ A {\bf 19}, 627 (2004)
  [hep-th/0310045].
  G.~J.~Olmo,
  Phys.\ Rev.\ D {\bf 72}, 083505 (2005)
  [gr-qc/0505135].
  S.~Baghram, M.~Farhang and S.~Rahvar,
  Phys.\ Rev.\ D {\bf 75}, 044024 (2007)
  [astro-ph/0701013].
  M.~S.~Movahed, S.~Baghram and S.~Rahvar,
  Phys.\ Rev.\ D {\bf 76}, 044008 (2007)
  [arXiv:0705.0889 [astro-ph]].
\bibitem{Hagihara:1931}
Y. ~Hagihara,
 Japan.\ J. Astron.\ Geophys, 8,67, (1931).
\bibitem{Hackmann:2010zz}
  E.~Hackmann, C.~Lammerzahl, V.~Kagramanova and J.~Kunz,
  Phys.\ Rev.\ D {\bf 81}, 044020 (2010)
  [arXiv:1009.6117 [gr-qc]].
\bibitem{Hackmann:2008tu}
  E.~Hackmann, V.~Kagramanova, J.~Kunz and C.~Lammerzahl,
  Phys.\ Rev.\ D {\bf 78}, 124018 (2008)
  [Phys.\ Rev.\  {\bf 79}, 029901 (2009)]
  [arXiv:0812.2428 [gr-qc]].
\bibitem{Enolski:2010if}
 V.~Z.~Enolski, E.~Hackmann, V.~Kagramanova, J.~Kunz and C.~Lammerzahl,
  J.\ Geom.\ Phys.\  {\bf 61}, 899 (2011)
  [arXiv:1011.6459 [gr-qc]].
\bibitem{Kagramanova:2012hw} 
  V.~Kagramanova and S.~Reimers,
  Phys.\ Rev.\ D {\bf 86}, 084029 (2012);
  V.~Diemer, J.~Kunz, C.~L\"ammerzahl and S.~Reimers,
  Phys.\ Rev.\ D {\bf 89}, no. 12, 124026 (2014)
\bibitem{Hackmann:2009rp}
E.~Hackmann, B.~Hartmann, C.~Laemmerzahl and P.~Sirimachan,
  Phys.\ Rev.\ D {\bf 81}, 064016 (2010)
  [arXiv:0912.2327 [gr-qc]].
\bibitem{Hackmann:2010ir}
  E.~Hackmann, B.~Hartmann, C.~Lammerzahl and P.~Sirimachan,
  Phys.\ Rev.\ D {\bf 82}, 044024 (2010)
  [arXiv:1006.1761 [gr-qc]].
\bibitem{Grunau:2012ai}
 S.~Grunau, V.~Kagramanova, J.~Kunz and C.~Lammerzahl,
  Phys.\ Rev.\ D {\bf 86}, 104002 (2012)
  [arXiv:1208.2548 [gr-qc]].
\bibitem{Grunau:2012ri}
  S.~Grunau, V.~Kagramanova and J.~Kunz,
  Phys.\ Rev.\ D {\bf 87}, no. 4, 044054 (2013)
  [arXiv:1212.0416 [gr-qc]].
\bibitem{Grunau:2013oca}
 S.~Grunau and B.~Khamesra,
  Phys.\ Rev.\ D {\bf 87}, no. 12, 124019 (2013)
  [arXiv:1303.6863 [gr-qc]].
\bibitem{Saffari:2007zt}
 R.~Saffari and S.~Rahvar,
  Phys.\ Rev.\ D {\bf 77}, 104028 (2008)
  [arXiv:0708.1482 [astro-ph]].
\bibitem{Abramowitz:1968}
M.~ Abramowitz and I.~ E. ~Stegun,
Dover Publications, New York,(1968).
\bibitem{Whittaker:1973}
E.~T.~ Whittaker and G.~N.~ Watson,
(reprinted 1973).
\bibitem{Buchstaber:1997}
V.~ M.~ Buchstaber, V.~ Z.~ Enolskii, and D.~ V.~ Leykin, 
 New York, 1997.
\bibitem{Gibbons:2011rh}
  G.~W.~Gibbons and M.~Vyska,
  Class.\ Quant.\ Grav.\  {\bf 29}, 065016 (2012)
  [arXiv:1110.6508 [gr-qc]].
\bibitem{Hartmann:2010rr}
 B.~Hartmann and P.~Sirimachan,
  JHEP {\bf 1008}, 110 (2010)
  [arXiv:1007.0863 [gr-qc]].
\end{thebibliography}

\end{document}